\documentclass[a4paper,11pt]{article}
\pdfoutput=1 

\usepackage{jheppub} 

\usepackage[T1]{fontenc} 

\usepackage{graphicx}
\usepackage{gensymb}
\usepackage{xcolor} 
\colorlet{lightgray}{black!10} 
\colorlet{pencil}{black!25} 
\usepackage{float}
\usepackage{cancel}
\usepackage{tkz-euclide} 
\usetikzlibrary{calc}
\usepackage{url}
\usepackage{pdflscape}
\usepackage{lscape}
\usepackage{dsfont}
\usepackage[utf8]{inputenc}
\usepackage{changepage}
\usepackage{caption}

\usepackage{float} 
\usepackage{amsmath}
\usepackage{mathtools}
\usepackage{amssymb}
\usepackage{amstext}
\usepackage{amsfonts}
\DeclareSymbolFontAlphabet{\amsmathbb}{AMSb}

\usepackage{bbold}
\usepackage{graphicx}

\usepackage{leftindex}

\usepackage{bbm}
\usepackage[many]{tcolorbox}

\newenvironment{nicebox}{}{}
\tcolorboxenvironment{nicebox}{
  enhanced,
  borderline={0.8pt}{0pt}{black},
  borderline={0.4pt}{2pt}{black},
  boxrule=0.4pt,
  colback=white,
  coltitle=black,
}

\newcommand{\AdS}{\text{AdS}}
\newcommand{\CFT}{\text{CFT}}
\newcommand{\AdSST}{\text{AdS}_5\times\text{S}^5}
\newcommand{\Sph}{\text{S}}

\newcommand{\psu}{\mathfrak{psu}}
\newcommand{\su}{\mathfrak{su}}

\newcommand{\SU}{\mathrm{SU}}
\newcommand{\U}{\mathrm{U}}
\newcommand{\PSU}{\mathrm{PSU}}
\newcommand{\SO}{\mathrm{SO}}

\newcommand{\ket}[1]{\left|{#1}\right\rangle}
\newcommand{\braket}[1]{\left\langle{#1}\right\rangle}

\newcommand{\h}{\mathbf{h}}
\newcommand{\tr}{\mathrm{tr}}

\newcommand{\ie}{\textit{i.e.}~}
\newcommand{\cf}{\textit{c.f.}~}
\newcommand{\eg}{\textit{e.g.}}

\title{Three-point functions from integrability \\ in $\mathcal{N}=2$ orbifold theories}

\author[a]{Dennis le Plat,}
\author[b]{Torben Skrzypek}

\affiliation[a]{HUN-REN Wigner Research Centre for Physics , Konkoly-Thege Mikl\'os u. 29-33, 1121 Budapest,
Hungary}
\affiliation[b]{Deutsches Elektronen-Synchrotron DESY, Notkestra{\ss}e 85, 22607 Hamburg, Germany}

\emailAdd{dennis.leplat@wigner.hu}
\emailAdd{torben.skrzypek@desy.de}

\abstract{Besides solving the spectral problem of $\mathcal{N}=4$ Super-Yang-Mills (SYM) theory, integrability also provides us with tools to compute the structure constants of the theory, most prominently through the hexagon formalism. We show that, with minor modifications, this formalism can also be applied to orbifolds of $\mathcal{N}=4$ SYM theory, which are integrable theories in their own right.
To substantiate this claim, we test our results against a direct gauge-theory calculation at tree-level. We focus here on a family of $\mathcal{N}=2$ supersymmetric $\amsmathbb{Z}_M$-orbifold theories. BPS correlators in these theories have recently been investigated with independent localisation techniques and a structural matching with wrapping corrections in the hexagon formalism was observed. Together with our weak-coupling evidence, this suggests that a full determination of the structure constants of orbifold theories at finite coupling may be within reach.}

\preprint{DESY-25-088}

\begin{document} 
\bibliographystyle{JHEP}
\maketitle
\flushbottom

\newpage
\section{Introduction} \label{sec:Introduction}
The $\AdS/\CFT$ correspondence is best understood for type IIB superstrings on $\AdSST$ and $\mathcal{N}=4$ supersymmetric Yang-Mills (SYM) theory \cite{Maldacena:1997re,Witten:1998qj,Gubser:1998bc}. It was noticed by Minahan and Zarembo \cite{Minahan:2002ve} that in the planar limit \cite{HOOFT1974461} the one-loop spectrum of anomalous dimensions of certain operators can be mapped to an integrable spin-chain Hamiltonian.
This allows to solve the spectrum exactly by using Bethe Ansatz techniques \cite{Bethe:1931hc}.
The map to integrable spin chains was then extended to the full superconformal algebra of $\mathcal{N}= 4$ SYM theory \cite{Beisert:2003yb,Beisert:2003jj,Beisert:2003tq}.
Considering operators of finite length, wrapping corrections have to be taken into account \cite{Bajnok:2008bm}. These finite-size corrections can be described on the string worldsheet by introducing a mirror model \cite{Arutyunov:2007tc,Bombardelli:2009ns,Gromov:2009bc,Arutyunov:2009ur}.
Thus, in principle the spectral problem of anomalous dimensions in $\mathcal{N}= 4$ SYM is solved and large amounts of data can be produced by virtue of the quantum spectral curve formalism \cite{Gromov:2013pga,Gromov:2014caa,Marboe:2017dmb,Gromov:2023hzc}.

Following the success of integrability in the spectral problem, the study of three-point functions of non-protected operators was initiated in \cite{Escobedo:2010xs}, where correlators of three closed spin-chain states were calculated at tree level.
\begin{figure}[h]
    \centering
    \includegraphics[width = .8\textwidth]{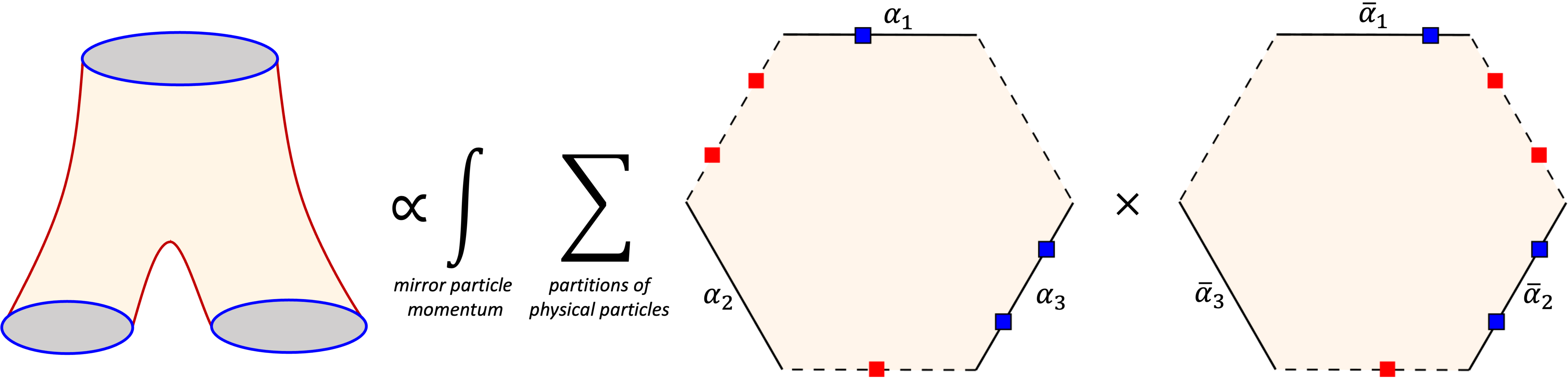}
    \caption{Structure constants can be evaluated through the hexagon formalism. After cutting the worldsheet, all possible distributions of excitations on the two hexagons have to be summed. Finite-size correction can be captured by inserting full sets of states on the cut edges.}
    \label{Fig:Hexagon}
\end{figure}
Subsequently, the hexagon formalism \cite{Basso:2015zoa} was developed to apply integrability techniques to three-point functions of generic operators in $\mathcal{N}=4$ SYM. Starting from the string theory picture, the worldsheet is cut into two hexagonal patches, \cf Fig. \ref{Fig:Hexagon}. Each patch has three edges corresponding to the cut operators (\textit{physical edges}) and three edges that correspond to the cut worldsheet (\textit{virtual edges}). The physical excitations carried by the operators can end up on either of the hexagons after cutting the operators and hence it is necessary to sum over all possible partitions of excitations. The hexagonal patches can be evaluated as form factors. Using the symmetries of three-point functions, the one- and two-particle hexagon form factor were bootstrapped and the multi-particle form factor was conjectured \cite{Basso:2015zoa}. Asymptotically, the bootstrap yields the $\psu(2|2)$ S-matrix elements \cite{Beisert:2005tm}. More specifically, the worldsheet is considered asymptotically large after cutting and hence finite-size corrections are suppressed.
In analogy to L\"uscher-corrections in field theory \cite{Luscher:1986pf}, finite-size effects can be captured order by order through the insertion of full sets of mirror particles on the virtual edges \cite{Basso:2015zoa}. Since this is relevant when gluing the two hexagonal patches back together into a three-point function, it is also referred to as gluing corrections. However, the explicit evaluation of these processes is rather involved, see for instance refs. \cite{Eden:2015ija,Basso:2015eqa,Basso:2017muf,Fleury:2016ykk,Fleury:2017eph,DeLeeuw:2019dak}.
Moreover, it is possible to extend the formalism to planar higher-point correlation functions \cite{Eden:2016xvg,Fleury:2016ykk}, and even to non-planar correlators \cite{Eden:2017ozn,Bargheer:2017nne,Bargheer:2019kxb}.

It seems natural to ask whether this very
promising program can be extended to theories with less supersymmetry. One possibility is to consider other instances of $\AdS/\CFT$ and in fact a first affirmative example was given for $\AdS_3 \times \mathrm{S}^3 \times \mathrm{T}^4$ \cite{Eden:2021xhe,Fabri:2022aup}.
Another possibility is to consider deformations of $\mathcal{N}=4$ SYM. For certain operators in the $\beta$- and $\gamma$-deformed theory a similar formalism seems applicable \cite{Eden:2022ipm} and it would be desirable to have a first-principles derivation.
For orbifold theories \cite{Kachru:1998ys,Lawrence:1998ja}, recent progress was made in \cite{Ferrando:2025qkr}, where a three-point function of BPS operators was considered and agreement with results from localisation \cite{Billo:2022xas, Billo:2022gmq, Billo:2022fnb} was found. The main goal of this article is to explore the hexagon formalism for orbifold theories for non-BPS operators.

\begin{figure}[h]
    \centering
    \includegraphics[width = .25\textwidth]{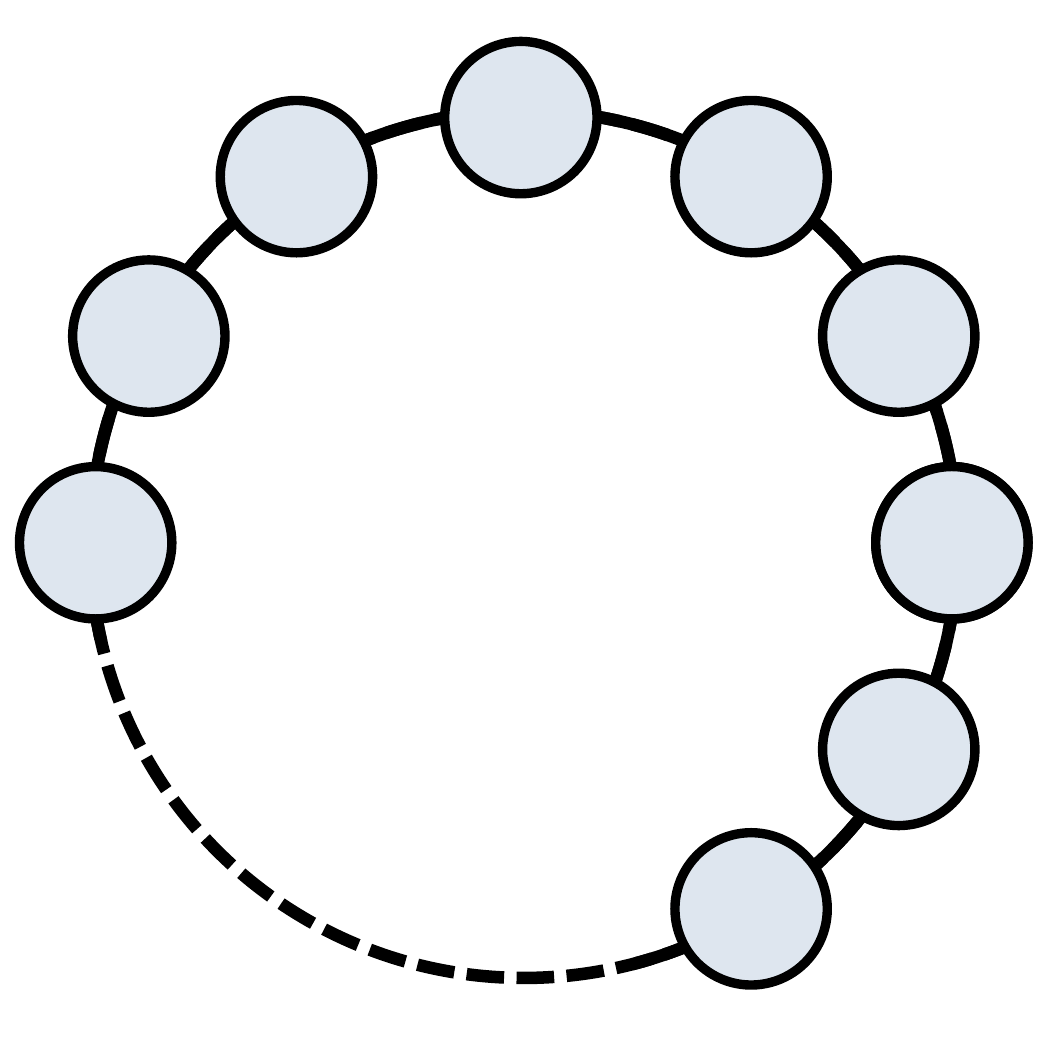}
    \caption{The ``necklace" quiver diagrams associated to $\mathcal{N}=2$ $\amsmathbb{Z}_M$-orbifold theories feature $M$ gauge nodes and bifundamental hypermultiplets. All gauge nodes have the same gauge coupling constant $g$.}
    \label{Fig:Quiver}
\end{figure}

Our main focus will be on $\mathcal{N}=2$ $\amsmathbb{Z}_M$-orbifolds of $\AdSST$ which arise as the near-horizon limit of a stack of D3-branes probing a $\amsmathbb{C}^2/\amsmathbb{Z}_M$ singularity ($A_{M-1}$ in the usual ADE-classification  \cite{Douglas:1996sw}). The dual $\mathcal{N}=2$ gauge theory consists of $M$ gauge multiplets and bifundamental hypermultiplets summarised by the quiver diagram Fig. \ref{Fig:Quiver}.
These orbifold theories are integrable \cite{Ideguchi:2004wm,Beisert:2005he,Skrzypek:2022cgg} and have also been investigated with localisation techniques, which allow us to calculate correlators of BPS-operators at all values of the gauge coupling~$g$~\cite{Pestun:2007rz, Pini:2017ouj,Galvagno:2020cgq,Billo:2022xas,Billo:2022gmq,Billo:2022fnb,Billo:2022lrv,Skrzypek:2023fkr}. However, the analysis of unprotected quantities will require the application of less specialised tools such as integrability. 

\begin{figure}[h]
    
    \centering 

\includegraphics[width = .33\textwidth]{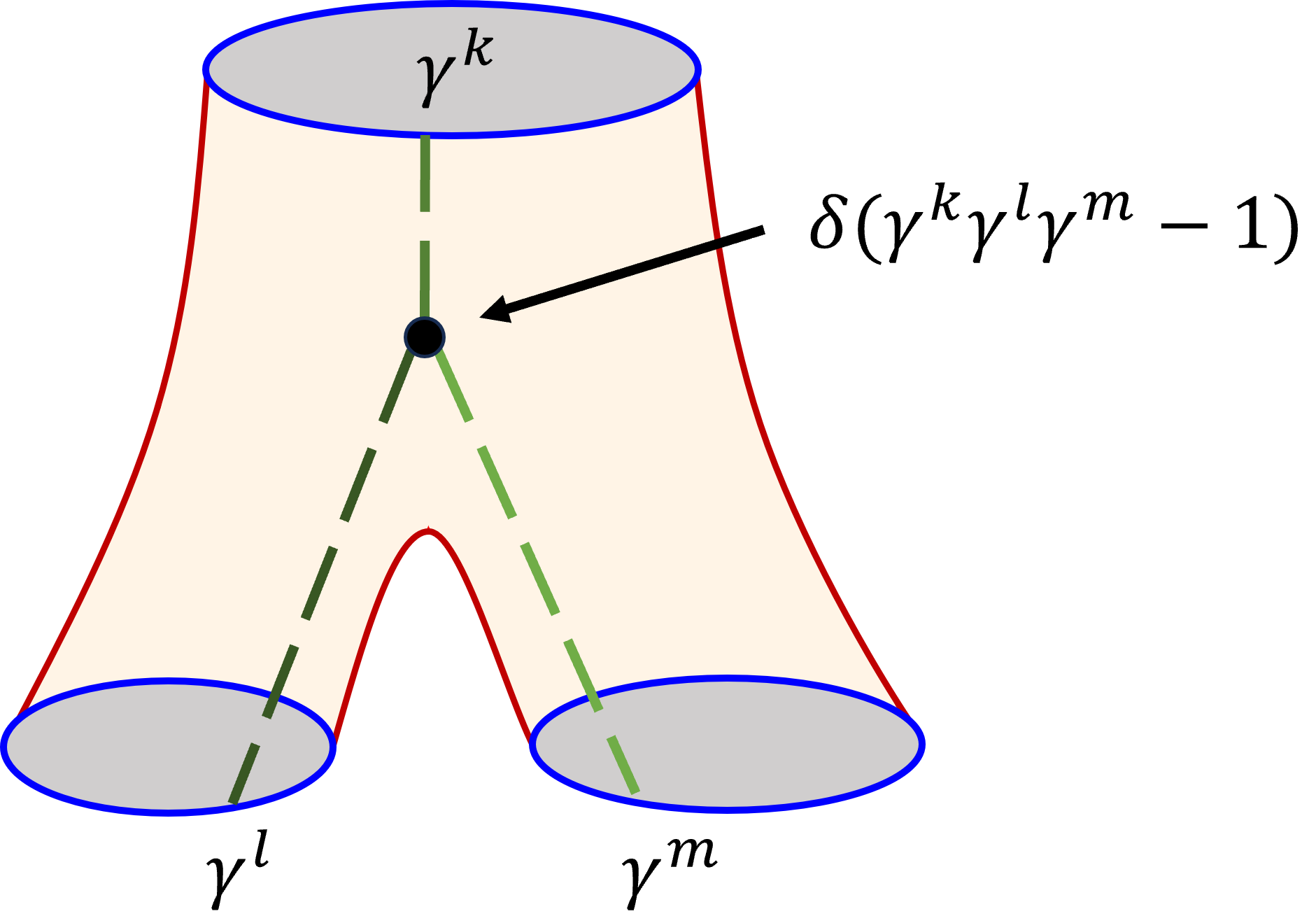}
    \caption{The twist operators inserted in the traces correspond to modified boundary conditions of the dual closed string states, which may be represented by twist-lines on the worldsheet \cite{Cavaglia:2020hdb}. The overall twist within a consistent three-point function has to cancel, resulting in an effective vertex for the twist lines that ensures this cancellation. Alternatively, one may split e.g. $\gamma^m=\gamma^{-k}\gamma^{-l}$ on the boundary and directly connect the twist lines from boundary to boundary. }
\label{Fig:Twistvertex}
\end{figure}

Along with changing the gauge structure of the fundamental fields, the orbifold projection also modifies the spectrum of single-trace operators by allowing for twisted-sector states that introduce an element $\gamma^k$ of the $\amsmathbb{Z}_M$-representation into the trace. These states are dual to string states that only close up to an orbifold action. When considering three-point functions of single-trace operators we therefore have to specify which twisted sectors we are contracting. Denoting scalar operators twisted by $\gamma^k$ as $\mathcal{O}^k$, overall orbifold invariance results in a superselection rule and conformal invariance fixes the space-time dependence
\begin{equation}\label{eq:superselection}
    \braket{\mathcal{O}^k(x_1)\mathcal{O}^l(x_2)\mathcal{O}^m(x_3)}=\frac{\mathcal{C}_{\mathcal{O}^k\mathcal{O}^l\mathcal{O}^m}\delta_{(k+l+m)\text{ mod } M}}{\vert x_1-x_2\vert^{\Delta_1+\Delta_2-\Delta_3}\vert x_1-x_3\vert ^{\Delta_1+\Delta_3-\Delta_2}\vert x_2-x_3\vert^{\Delta_2+\Delta_3-\Delta_1}}\,,
\end{equation}
where we denote by $\Delta_i$ the conformal dimension of the operator inserted at $x_i$. The superselection rule may be understood pictorially by allowing the three twist lines to merge on the dual string worldsheet, see Fig. \ref{Fig:Twistvertex}. 

Our goal will be to determine the value of the structure constants $\mathcal{C}_{\mathcal{O}^k\mathcal{O}^l\mathcal{O}^m}$ by making use of the hexagon formalism outlined above. These structure constants depend on operator normalisations, which we may mitigate by considering ratios of the three-point functions of interest with a reference three-point function. As reference we choose the protected three-point function involving only untwisted BPS-operators $\mathcal{V}^0$ of the same lengths, resulting in the ``normalised" three-point function\footnote{This choice assumes a uniform normalisation scheme across twist sectors, as e.g. used in \cite{Skrzypek:2023fkr}.}
\begin{equation}\label{eq:3-pt}
    \frac{\braket{\mathcal{O}^k(x_1)\mathcal{O}^l(x_2)\mathcal{O}^m(x_3)}}{\braket{\mathcal{V}^0(x_1)\mathcal{V}^0(x_2)\mathcal{V}^0(x_3)}}=\frac{\hat{\mathcal{C}}_{\mathcal{O}^k\mathcal{O}^l\mathcal{O}^m}\delta_{(k+l+m)\text{ mod } M}}{\vert x_1-x_2\vert^{\gamma_1+\gamma_2-\gamma_3}\vert x_1-x_3\vert ^{\gamma_1+\gamma_3-\gamma_2}\vert x_2-x_3\vert^{\gamma_2+\gamma_3-\gamma_1}}\,,
\end{equation}
which now only depends on the anomalous dimensions $\gamma_i=\Delta_i-L_i$. In an abuse of notation we will sometimes denote the structure constants by their associated correlators $\hat{\mathcal{C}}_{\mathcal{O}^k\mathcal{O}^l\mathcal{O}^m}\hat{=}\braket{\mathcal{O}^k\mathcal{O}^l\mathcal{O}^m}$, suggesting that the spacetime-dependence and normalisation can be restored by reference to \eqref{eq:3-pt}. It is these normalised structure constants that the hexagon formalism will determine. In order to check our results, we compare them with a direct gauge-theory calculation at tree level. 

The set-up of this calculation is complicated by the reduced amount of symmetry in the orbifold theory. In $\mathcal{N}=4$ SYM one usually prepares the operators in question at one point in spacetime and translates them to a finite distance using a $\PSU(2|2)$-preserving translation operator $\mathcal{T}$. This operator mixes the original state with other operators from the same superconformal multiplet. The three-point function then picks out the component that is appropriately rotated to preserve overall $R$-symmetry and we may determine the associated structure constant via the hexagon formalism. In the $\mathcal{N}=2$ orbifold the translation operator $\mathcal{T}$ generally does not commute with the orbifold twist and the overall symmetry is broken to at best $\PSU(1|1)$. This results in the mixing of several superconformal multiplets with distinct anomalous dimensions and structure constants. Nevertheless, we find evidence that a minor modification of the usual $\mathcal{N}=4$ hexagon formula still determines individual structure constants accurately and matches the gauge-theory prediction after the latter has been projected onto the contributions of individual superconformal multiplets. We therefore conjecture the following statement:

\begin{nicebox}
The hexagon formalism accurately determines the structure constants of physical states in $\mathcal{N}=2$ $\amsmathbb{Z}_M$-orbifold theories, if:
\begin{itemize}
\item we consider solutions to the twisted Bethe equations and 
\item keep track of the corresponding twist factors for particles being moved past the twist operators during the cutting and gluing procedure. 
\end{itemize}
\end{nicebox}

In this paper we will provide tree-level evidence for this statement at low numbers of magnon excitations in various $\SU(2)$ sectors of the simplest $\amsmathbb{Z}_2$ orbifold as well as some preliminary checks for the $\amsmathbb{Z}_3$ orbifold, which directly generalise to $\amsmathbb{Z}_M$ orbifolds. A recent discussion of wrapping corrections \cite{Ferrando:2025qkr} provided additional evidence that a correct treatment of mirror excitations follows the same logic by matching to localisation results. In light of these promising results, a full determination of the structure constants of non-BPS operators at finite coupling seems possible and would be an interesting challenge for future research.

We start our discussion by reviewing the spectral problem of orbifold theories in Section \ref{sec:spectral}. We first introduce the  $\mathcal{N}=2$ $\amsmathbb{Z}_M$-orbifolds and their gauge theory incarnation (\cf Fig. \ref{Fig:Quiver}), comment on the gauge choice necessary to set up a spin-chain picture and finally present some simple $\SU(2)$ sectors of single-trace operators. An explicit calculation of their spectrum is performed for the $\amsmathbb{Z}_2$-orbifold theory and we listed states with up to two excitations in Appendix \ref{app:Collection}. In Section \ref{Sec:ThreeApproaches} we  briefly review the available technology to compute three-point functions, directly from gauge theory, via a spin-chain overlap and through the hexagon formalism. In Section \ref{Sec:Z2Hexagon} we apply these techniques to the orbifold theories. We first compute three-point functions for the simplest gauge choice in the $\amsmathbb{Z}_2$ orbifold where an additional $\SU(2)$-symmetry appears. This symmetry mitigates the need to project onto the relevant superconformal multiplets as their structure constants are now related by this symmetry. When choosing a different gauge or moving on to higher-order $\amsmathbb{Z}_M$ orbifolds this quality is lost and we need to decompose the three-point functions that arise from the translated operators into individual components. In general, this would be a very tedious endeavour but for some simple operators we can indeed show agreement with the hexagon. It stands to reason that we may therefore take the hexagon as a direct access route to structure constants. We comment on possible additional checks, extensions to finite coupling and other deformations of $\mathcal{N}=4$ SYM in Section \ref{Sec:Comclusion}.

\section{Review: The spectral problem of orbifold theories}\label{sec:spectral}

In this Section we will gather the necessary background on orbifold theories, their symmetries and the associated spin-chain. A simple subset of operators is given by $\SU(2)$ sectors which consist of single-trace operators involving only two scalar fields. In orbifold theories we can construct multiple $\SU(2)$ sectors which are distinguished by their transformation properties under the orbifold action. We will list the various $\SU(2)$ sectors of $\mathcal{N}=2$ supersymmetric orbifold theories.  

\subsection{Orbifolds theories}

The simplest integrable deformation of type IIB string theory on $\AdSST$ is an orbifold by some discrete subgroup $\Gamma$ of the overall $\PSU(2,2\vert 4)$ symmetry \cite{Kachru:1998ys,Lawrence:1998ja}. The closed-string spectrum of such orbifold theories is constructed by projecting the undeformed spectrum onto $\Gamma$-invariant states (untwisted sector) and furthermore adding additional string states which close only up to a $\Gamma$-action and were absent in the undeformed theory (twisted sectors). We thus have to combine $\vert\Gamma\vert$ different sectors to achieve a modular invariant string theory. In this paper, we will restrict our attention to cyclic groups $\Gamma=\amsmathbb{Z}_M$ acting on the $\Sph^5$ subspace, or in terms of the dual gauge theory on the $R$-symmetry subgroup $\SU(4)$.

The dual gauge theory descends from $\mathcal{N}=4$ SYM  theory with gauge group $\SU(MN)$ on which the orbifold group $\Gamma=\amsmathbb{Z}_M$ acts as \cite{Ideguchi:2004wm,Beisert:2005he}
\begin{equation}
    \gamma=\text{diag}(\mathbb{1}_{N},\omega\cdot\,\mathbb{1}_{N},\omega^2\cdot\,\mathbb{1}_{N}\,,\dots\,,\omega^{M-1}\cdot\,\mathbb{1}_{N})\,,\qquad \omega=\exp{\frac{2\pi i}{M}}\,,
\end{equation}
and thus breaks the gauge group down to $\U(N)^M$\,. We furthermore need to act with the appropriate $R$-symmetry transformation $R_\gamma$ on the various fields in the theory. Imposing invariance under the combined action of $\gamma$ and $R_\gamma$ will generally break supersymmetry completely. If we want to retain supersymmetry, we have to ask for $R_\gamma$ to leave an appropriate amount of $R$-symmetry unbroken, as is best illustrated by the action on the $\mathcal{N}=4$ scalar fields. If $\mathcal{N}=1$ supersymmetry shall be preserved, we may parametrise the most general orbifold action as 
\begin{equation}
R_\gamma^{-1}(X,Y,Z)=\gamma ^\dagger (X,Y,Z)\gamma=(\omega^{t_X} X,\omega^{t_Y} Y,\omega^{t_Z} Z)\,,\qquad t_X,t_Y,t_Z\in \amsmathbb{Z} \text{ mod } M\,, \label{eq:GeneralOrbiAction}
\end{equation}
preserving the $\U(1)$ $R$-symmetry under which the complexified scalars $X,Y$ and $Z$ are charged. If we want to retain $\mathcal{N}=2$ supersymmetry the orbifold action may be restricted to
\begin{equation}\label{eq:orbifold action}
 R_\gamma^{-1}(X,Y,Z)=\gamma ^\dagger (X,Y,Z)\gamma=(\omega X,\omega^{-1} Y, Z)\,,
\end{equation}
which furthermore preserves an $\SU(2)_R\subset \SO(4)$ $R$-symmetry. The simplest case of this orbifold action is the $\amsmathbb{Z}_2$ orbifold for which the twist factor $\omega=\omega^{-1}=-1$ degenerates and allows for yet another $\SU(2)$-symmetry between $X$ and $Y$. In Section \ref{ssec:Simplesec} we will study three-point functions in this theory and see that this additional $\SU(2)$-symmetry greatly simplifies the calculations. 

Let us take a closer look at the $\mathcal{N}=2$ orbifold, following in part the discussion in \cite{Ideguchi:2004wm}.  Invariance under the combined gauge and $R$-symmetry transformations according to \eqref{eq:orbifold action} dictates that the $\SU(MN)$-adjoint fields $X,Y$ and $Z$ decompose into bifundamental and adjoint fields of $\U(N)^M$ as follows
\begin{equation}
X = \begin{pmatrix}
0 & X_{12} & \\
& 0& X_{23} & \\
&  & \ddots &\ddots \\
X_{M1} & & & 0
\end{pmatrix}\,, \quad 
Y = \begin{pmatrix}
0 & & & Y_{1M} \\
Y_{21} & 0 & & \\
& Y_{32} & \ddots & \\
& & \ddots & 0
\end{pmatrix}\,,\quad
Z = \begin{pmatrix}
Z_{11} & & \\
& Z_{22} & \\
& & \ddots & \\
& & & Z_{MM}
\end{pmatrix}\,,\label{eq:GaugeMatrix}
\end{equation}
where the indices label the respective gauge groups. This break-down of the $\SU(MN)$ gauge group now naturally encodes the projection to $\amsmathbb{Z}_M$-invariant states in the dual string picture. The holographic dictionary maps closed string states to single-trace operators
\begin{equation}
    \tr_{MN} ZXXYZ\bar{Z}\dots\,.
\end{equation}
The trace becomes a sum over traces $\tr_N$ of the individual $\U(N)$-factors, but only if the first and last gauge indices match. This projects out all states that do not satisfy the orbifold condition
\begin{equation}\label{eq:winding}
 \#(X)-\#(Y)-\#(\bar{X})+\#(\bar{Y})=w M \,, \qquad w\in\amsmathbb{Z}\,,
\end{equation}
and precisely the untwisted sector states remain. Here we defined a ``winding number" $w$ which counts the number of times the state wraps around the quiver diagram Fig. \ref{Fig:Quiver}. 
We may similarly construct the twisted sector states by adding powers of $\gamma$ to the trace
\begin{equation}
    \tr_{MN} \gamma^k ZXXYZ\bar{Z}\dots\,,
\end{equation}
which translates the twisted boundary conditions of the string to the gauge theory and introduces relative phases into the sum of $\U(N)$-traces.

In \cite{Bertle:2024djm} it was pointed out that although the orbifold action breaks the $\SU(4)$ $R$-symmetry, one may restore the broken elements of the Lie algebra in the language of algebroids. This captures the fact that acting with broken generators $R_b$ on fields would change their gauge indices from one gauge group to another, e.g. 
\begin{equation}
    R_b:\quad X_{12}\to Z_{11}\,.
\end{equation}
To reinstate such algebroid elements to the theory, one may introduce an algebroid co-product
\begin{equation}
\Delta(R_b)=\mathbb{1}\otimes R_b+R_b\otimes \Omega\,, \label{eq:AlgebroidCoproduct}
\end{equation}
where $\Omega$ acts as a $\amsmathbb{Z}_k$-rotation on the  gauge indices. Essentially, the prescription is to rotate the gauge indices to the right of a transformed operator in such a way as to allow contractions with the new gauge index. Within traces, the notion of ``to the right" is not well-defined so the authors of \cite{Bertle:2024djm} proposed an opening procedure that temporarily allows for open spin chains as intermediate states. Repeated action of broken generators may eventually lead back to a non-vanishing trace. We will make use of this language when discussing the interplay of translations with the orbifold action.

\subsection{Gauge choice and symmetries}\label{ssec:gauge}

For sufficiently long single-trace operators, the spectral problem of $\mathcal{N}=4$ SYM has been solved by the asymptotic Bethe Ansatz. Its starting point is a BPS-vacuum built from a scalar operator $\Phi_V$ (a canonical choice is $\Phi_V=Z$, but any $\SO(6)$-rotation thereof is possible)   
\begin{equation}\label{eq:vacuum}
    \tr\, \Phi_V^L\,,
\end{equation}
which is interpreted as an integrable spin chain of length $L$ into which further operators, such as other scalars or derivatives, are inserted as quasi-particle excitations. The choice of the vacuum corresponds to choosing a light-cone gauge in the dual string theory and breaks the overall supersymmetry from $\PSU(2,2|4)$ down to $\PSU(2|2)^2$.  To make this breaking explicit, consider the fundamental representation of $\psu(2,2|4)$ in terms of complex 8-by-8 matrices. A generic generator may be block-decomposed as
\begin{equation}\label{eq:matrixrep}
    \left(
    \begin{array}{c c|c c}
      L^\alpha{}_\beta + D & K^\alpha{}_{\dot{\beta}} & \bar{Q}^\alpha{}_b& \bar{Q}^\alpha{}_{\dot{b}}\\
      P^{\dot{\alpha}}{}_{\beta} & L^{\dot{\alpha}}{}_{\dot{\beta}} - D &  Q^{\dot{\alpha}}{}_b& Q^{\dot{\alpha}}{}_{\dot{b}}\\
      \hline
        \bar{S}^a{}_\beta & S^a{} _{\dot{\beta}}  & R^a{}_b -J& R^a{}_{\dot{b}}\\
      \bar{S}^{\dot{a}}{}_\beta & S^{\dot{a}}{}_{\dot{\beta}}  & R^{\dot{a}}{}_b& R^{\dot{a}}{}_{\dot{b}}+J\\
    \end{array}
    \right)\,,
\end{equation}
where we introduced fundamental $\SU(2)$ indices $\alpha,\dot{\alpha},a,\dot{a}\in\{1,2\}$ which determine the transformation behaviour under the $\SU(2)$ subgroups generated by $ L^\alpha{}_\beta$, $L^{\dot{\alpha}}{}_{\dot{\beta}}$, $R^a{}_b$ and $ R^{\dot{a}}{}_{\dot{b}}$, respectively.
While the decomposition of the conformal symmetry group $\SU(2,2)$ (in the upper-left quadrant) according to the maximal compact subgroup $\SU(2)\times \SU(2)$ is natural, the decomposition of the $\SU(4)$ $R$-symmetry group (in the lower-right quadrant) may seem ad-hoc and in fact one may choose any $\SU(4)$ conjugated version as an equivalent decomposition. We shall choose the orientation such that the vacuum \eqref{eq:vacuum} has charge $L$ under the action of the generator $J$. Since this operator \eqref{eq:vacuum} is BPS, its eigenvalue under the dilatation operator $D$ is fixed to $\Delta=J$. Thus, we may interpret the vacuum \eqref{eq:vacuum} as the ground-state w.r.t. to the light-cone Hamiltonian 
\begin{equation}\label{eq:Hamiltonian}
    \mathcal{H}=D-J\,,
\end{equation}
which we interpret as the Hamiltonian of the associated spin-chain. 
A basis of excitations that diagonalises this Hamiltonian will furnish a representation of the commutant $C_{\PSU(2,2|4)}(\mathcal{H})$, which is spanned by the generators with purely dotted or undotted indices. These form two $\PSU(2|2)$ subgroups that can be centrally-extended to generate the full off-shell algebra of quasi-particles. The large amount of symmetry allows for a complete determination of the S-matrix of such excitations scattering off each other and we may employ a Bethe Ansatz to also determine the energies of multi-particle states \cite{Bethe:1931hc,Minahan:2002ve}. 

When applying this technology to orbifold theories, the various choices of the vacuum field $\Phi_V$ are no longer related by $\SO(6)$-symmetry. Instead they are now distinguished by their charge under the orbifold action \eqref{eq:orbifold action}. We may identify two main gauge choices: 
\newpage
\begin{itemize}
    \item \textbf{Adjoint vacuum}: $\Phi_V=Z$

    This vacuum choice is natural when comparing to $\mathcal{N}=4$ SYM. The vacuum field $Z$ sits in the adjoint representation of the $\mathcal{N}=2$ gauge theory and all (untwisted and twisted sector) single-trace operators built from it are BPS-operators. The winding number \eqref{eq:winding} of this vacuum is $w=0$. The orbifold action on the scalar fields was given in \eqref{eq:orbifold action} but after the gauge choice, we may now capture it in the language of the fundamental $\SU(4)$ representation (lower-right quadrant of \eqref{eq:matrixrep})
    \begin{equation}
        \label{eq:adjoint}R_\gamma=\begin{pmatrix}
            \omega & 0 &0 &0\\
            0 & \omega^{-1} & 0 & 0\\
            0 &0&1&0\\
            0&0&0&1
        \end{pmatrix}\,.
    \end{equation}
    It is thus an element of the $\SU(2)\subset\PSU(2|2)$ spanned by the undotted elements $R^a{}_b$. The other $\PSU(2|2)$ group remains unbroken by the orbifold projection.
    
    \item \textbf{Bifundamental vacuum} $\Phi_V=X$
    
    In this case, the vacuum field is charged under the orbifold action.
    When it comes to constructing single-trace operators according to \eqref{eq:vacuum}, only  untwisted vacua of the form $\tr(\gamma^0 X^{Mw})$ with positive winding number $w$ are part of the spectrum.\footnote{We could have equally chosen $Y$ or a conjugate fields $\bar{X}$, $\bar{Y}$ as vacuum field. The $Y$- and $\bar{X}$-vacuum states have negative winding numbers $w$.} In the twisted sector and for length $L\neq0\text{ mod } M$, we require at least one excitation to build a physical state.     In this gauge the orbifold action is given by 
    \begin{equation}
        \label{eq:bifundamental}R_\gamma=\begin{pmatrix}
           \omega & 0 &0 &0\\
            0 & 1 & 0 & 0\\
            0 &0&\omega^{-1}&0\\
            0&0&0&1
        \end{pmatrix}\,,
    \end{equation} 
    which breaks the supersymmetry down to $\PSU(1|1)^2$.
\end{itemize}
We could investigate further more complicated gauge choices, but since we have covered all winding numbers $w$, these would not offer a qualitatively different or more convenient scenario.

\subsection{$\SU(2)$ sectors} \label{sec:Review}

An analysis of the full spectrum of operators and their structure constants is the declared goal of the integrability program. In this paper, however, we will restrict our attention to certain subsectors of states in which the difference of $\mathcal{N}=2$ orbifold theories and $\mathcal{N}=4$ SYM become apparent, leaving a more extensive survey of structure constants to future work. The sectors in question consist of single-trace operators 
\begin{equation}\label{eq:SU(2)}
    \tr\,\gamma^k\Phi_V^{L-K}\Phi_E^{K}+\text{permutations}\,,
\end{equation}
involving only two scalar fields, the vacuum field $\Phi_V$ and the excitation field $\Phi_E$. We can go to the spin-chain picture by formally identifying $\Phi_V$ as down- ($\ket{\downarrow}$) and $\Phi_E$ as up-state $\ket{\uparrow}$. Hence single-trace operators can be identified with spin chain states as
\begin{equation}\label{eq:SU(2)R}
	\tr \gamma^k \Phi_V \Phi_E \Phi_V\Phi_V \Phi_E\dots  \quad \leftrightarrow \quad \ket{\downarrow \uparrow \downarrow \downarrow \uparrow \dots }^k \,,
\end{equation}
where the boundary twist has to be taken into account whenever the spin chain is shifted by one site. The eigenstates of the light-cone Hamiltonian \eqref{eq:Hamiltonian} can be obtained through diagonalising the one-loop dilatation operator $\mathcal{D}_2 = \sum_{l=1}^L \mathbb{1} - \amsmathbb{P}_{l,l+1}$, where $\amsmathbb{P}_{l,l+1}$ is the permutation operator acting on neighbouring fields. The corresponding eigenvalues are non-trivial corrections to the conformal scaling dimensions of the operators \eqref{eq:SU(2)}, which can be calculated via a Bethe Ansatz.  
Assuming a generic behaviour under the orbifold action \eqref{eq:orbifold action} as $(\Phi_V,\,\Phi_E) \sim (\omega^p \Phi_V,\, \omega^q \Phi_E)$, the Bethe equations are given by \cite{Ideguchi:2004wm}
\begin{equation}\label{eq:BAE}
	e^{i p_j L}  \prod_{j=1,j\neq i} ^{K}  S_{j,i}  = \omega^{k(p-q)} \,, \qquad
	e^{iP} = \prod_{j=1}^{K} e^{i p_j}  = \omega^{kp} \,, 
\end{equation}
where we introduced the S~matrix~$S_{j,i}$ scattering the excitations with momenta $p_j$ and $p_i$ as well as the shift factors~$e^{i p}$. In terms of rapidities and at leading order in the gauge coupling $g$ these are given by 
\begin{equation}
	S_{j,k} = \frac{u_j-u_k-i}{u_j-u_k+i}\,, \qquad e^{i p_j} =\frac{u_j +i/2}{u_j - i/2}\,. \label{eq:Smatrix}
\end{equation}

We will solve these equations for up to two excitations in the various scenarios below, but the generalisation to higher excited states is a matter of stamina. After determining the Bethe roots $u_i$, we can simply add the energies of the excitations to find the overall energy of the state
\begin{equation}
	E = \sum_{j=1}^{K} \frac{1}{u_i^2 +1/4} \,. \label{eq:AnomDim}
\end{equation}
The Bethe eigenstate can be constructed as in the undeformed case
\begin{equation}
\ket{\Psi}^k =  \sum_{1 \leq n_1 < \dots < n_{K} \leq L} \, \sum_{\sigma \in S_{K}} e^{i \sum_{l=1}^{K} p_{\sigma(k)} n_k} 
\prod_{\substack{j > l \\ \sigma(j) < \sigma(l)}} S_{{\sigma(j)}, {\sigma(l)}} \ket{n_1,\dots,n_{K}}^k \,, \label{eq:BetheState}
\end{equation}
where the sums run over all spin-chain sites as well as all permutations $\sigma$ of the $K$ excitations, which are inserted at the sites $n_j$ as indicated in the ket state. In particular, the twisted boundary conditions enter only implicitly through the Bethe roots $u_i$. The normalised state is then given by \cite{Korepin:1982ej,Gaudin:1996}
\begin{equation}
	\mathcal{B}^k = \frac{\ket{\Psi}^k}{\sqrt{\mathcal{G} \prod_j (u_j^2 + \frac{1}{4}) \prod_{i<j} S_{i,j}}} \,, \label{eq:BetheNorm}
\end{equation}
where the Gaudin norm $\mathcal{G}$ is defined as the logarithmic derivative of the Bethe equations, which is again independent of the twist factors. For the reader's convenience, we spell out the Gaudin norm in the $\SU(2)$ sectors, which explicitly reads
\begin{equation}
	\mathcal{G} = \mathrm{Det} \, \phi_{jl}\,, \qquad \phi_{jl} = -i \frac{\partial \log( e^{i p_j L} \prod_{j\neq l} S(u_j,u_l))}{\partial u_l} \,.
\end{equation}

We see that the only effect of the twist $\gamma^k$ is the introduction of phases $\omega^{k(p-q)}$ and $\omega^{kp}$ to the Bethe equations \eqref{eq:BAE}. It will therefore be useful to distinguish a few archetypical $\SU(2)$ sectors in the orbifold theory generated by \eqref{eq:orbifold action}, which we also list in Table \ref{tab:SU2Sectors}. We chose the bifundamental vacuum $\Phi_V=X$ here but one may always exchange vacuum and excitation fields at the cost of exchanging highly and slightly excited states.

\begin{table}
\centering
\begin{tabular}{c|c|c|c}
Vacuum & Sector & $(\Phi_V,\Phi_E)$ & $(p,q)$\\
\hline
bifundamental & $\SU(2)_{L}$ & $(X,\bar{Y})$ & $(1,1)$ \\
bifundamental & $\SU(2)_{R}$ & $(X,Y)$ & $(1,-1)$\\
bifundamental & mixed $\SU(2)$ & $(X,Z)$ & $(1,0)$ \\
adjoint & mixed $\SU(2)$ & $(Z,X)$ & $(0,1)$
\end{tabular}
\caption{A list of the different $\SU(2)$ sectors considered in the main text. All other $\SU(2)$ sectors can be related to these four cases by $R$-symmetry rotations and conjugations. 
The excitations $\Phi_E$ are inserted on a vacuum consisting of $\Phi_V$. Using the values $(p,q)$ in \eqref{eq:BAE} leads to the respective Bethe equations for the sector. Note that the $\SU(2)$ symmetry is broken for the mixed sector. } \label{tab:SU2Sectors}
\end{table}

\paragraph{$\SU(2)_R$ sector.} The orbifold action preserves an $\SU(2)_R$ symmetry between $X$ and $\bar{Y}$, and similarly for their complex conjugate fields. One may therefore study the scenario $\Phi_V=X$, $\Phi_E=\bar{Y}$ (or any $\SU(2)_R$-rotation thereof) and consider states of the form
\begin{equation}
	\tr \gamma^k \bar{Y}^{|J_2|} X^{J_1}  + \text{permutations}\,, \qquad(J_1>0\,, J_2<0)\,.
\end{equation}
These operators have the $\SO(6)$ charges $(J_1,J_2,0)$. The spectrum can be found from the Bethe Ansatz equations \eqref{eq:BAE} with $p=q=1$. The boundary condition is periodic and hence the Bethe equations are given by
\begin{equation}
	e^{i p_j L} \prod_{j=1,j\neq i} ^{|J_2|} S_{j,i} = 1 \,, \qquad e^{iP} = \prod_{j=1}^{|J_2|} e^{i p_j}= \omega^{k} \,, \label{eq:BetheEq}
\end{equation}
where the twist only appears in the constraint for the total momentum $P$.

\paragraph{${\SU(2)_L}$ sector.} We may also consider operators consisting of fields $X$ and $Y$. The $\SU(2)_L$-symmetry relating these fields is explicitly broken by the orbifold action \eqref{eq:orbifold action}. When considering operators of the form
\begin{equation}
	\tr \gamma^k {Y}^{J_2} X^{J_1}  + \text{permutations}\,, \qquad(J_1>0\,, J_2>0)\,,
\end{equation}
we therefore find a phase in the  Bethe equation for excitations
\begin{equation}
	e^{i p_j L} \prod_{j=1,j\neq i} ^{J_2} S_{j,i} = \omega^{2k} \,, \qquad e^{iP} = \prod_{j=1}^{J_2} e^{i p_j} = \omega^{k} \,.  \label{eq:BetheEqYB}
\end{equation}
Note that, in the special case of the $\amsmathbb{Z}_2$ orbifold, the Bethe equations of the $\SU(2)_R$ and the $\SU(2)_L$ sector coincide.

\paragraph{Mixed $\SU(2)$ sector.} Finally, we could also consider operators consisting of fields $X$ and $Z$ (or similarly $\bar{Z}$). In the original $\mathcal{N}=4$ SYM theory these fields also form an $\SU(2)$ sector. However, in the orbifold theory this symmetry is broken and now mixes adjoint and bifundamental representations. Like in the sectors considered above, single-trace operators take the form
\begin{equation}
	\tr \gamma^k Z^{J_3}{X}^{J_1}   + \text{permutations}\,, \qquad (J_1>0\,, J_3>0)\,.
\end{equation}
Building the state on top of the bifundamental vacuum the corresponding Bethe equations read
\begin{equation}
	\text{bifundamental vacuum:} \qquad e^{i p_j L} \prod_{j=1,j\neq i} ^{J_3} S_{j,i} = \omega^k \,, \qquad e^{iP} = \prod_{j=1}^{J_3} e^{i p_j} = \omega^k \,.  \label{eq:BetheEqX}
\end{equation}

For later convenience, let us also give the Bethe equations for excitations on top of the adjoint vacuum. In this case the Bethe equations read
\begin{equation}
	\text{adjoint vacuum:} \qquad e^{i p_j L} \prod_{j=1,j\neq i} ^{J_1} S_{j,i} = \omega^{-k} \,, \qquad e^{iP} = \prod_{j=1}^{J_1} e^{i p_j} = 1 \,.  \label{eq:BetheEqAdj}
\end{equation}
This list exhausts the physically distinguished $\SU(2)$ sectors up to duality transformations. 
As instructive example, let us analyse the spectrum of the $\mathcal{N}=2$ $\amsmathbb{Z}_2$-orbifold.

\subsection{Spectrum of the $\SU(2)$ sectors for the $\amsmathbb{Z}_2$ orbifold} \label{sec:SU2Sectors}
We will now construct some explicit states with two excitations for the $\amsmathbb{Z}_2$ orbifold. Since the untwisted sector descends directly from $\mathcal{N}=4$ SYM, we will instead focus on twisted-sector operators. For the $\amsmathbb{Z}_2$ orbifold the twist factor is simply $\omega=-1$, and therefore the $\SU(2)_R$ and the $\SU(2)_L$ sector are characterised by the same Bethe equation, \cf eqs. \eqref{eq:BetheEq} and \eqref{eq:BetheEqYB}.
Let us define a basis of doubly excited operators of length $L$ as
\begin{equation}
	\mathcal{O}_j^{k,L} = \tr (\gamma^k \Phi_E \Phi_V^j \Phi_E \Phi_V^{L-j-2} )\,. \label{eq:OpBasis}
\end{equation}
For each length of the operator we have a set of $2 \lfloor L/4 \rfloor$ distinct operators. Considering the mixing problem reveals a highly degenerate spectrum. The operator $\mathcal{O}^{1,L}_0$ has energy\footnote{This operator corresponds to a bound-state and is solved by the Bethe string $u_1=-u_2=\frac{i}{2}$. These operators exist in the $\SU(2)_{L,R}$ as well as the mixed $\SU(2)$ sector over the bifundamental vacuum. Although interesting in their own right, we disregard these operators when considering three-point functions, as the nature of these Bethe solutions complicates the evaluation.} $E=2$, while all the other operators $\mathcal{O}^{1,L}_j$, with $1\leq j \leq 2 \lfloor L/4 \rfloor$ have the energy $E=4$.

Even though the high degeneracy prevents us from finding good eigenstates from a naive diagonalisation of the one-loop Hamiltonian, the Bethe equations know about higher conserved charges present in an integrable theory. Choosing a basis of states according to the Bethe roots additionally diagonalises the state with respect to those.\footnote{We thank Paul Ryan for clarifying this point to us.} Solving the momentum constraint from \eqref{eq:BetheEq} for two excitations yields
\begin{equation}
	u_2 =  -\frac{1}{2 i} \frac{2 u_1 \left(\omega ^k+1\right) - i \left(\omega ^k-1\right)}{2 u_1 \left(\omega ^k-1\right) - i \left(\omega ^k+1\right)} \quad \Rightarrow \quad  u_2 = \frac{1}{4 u_1}\,,
\end{equation}
with $\omega=-1$ and $k=1$ for the twisted sector. We again observe a high degeneracy in this subsector, as the energy \eqref{eq:AnomDim} is $E=4$ for doubly excited states, irrespective of their length $L$.
Solving the Bethe equation, it turns out that primary operators with two excitations only exist for length $L \geq 8$. In Tab. \ref{tab:Z2Operators} the shortest possible primary operators are listed, as well as their eigenstates and rapidities.

We can proceed similarly for the mixed $\SU(2)$ sector. The spectrum of operators carrying two excitations is again highly degenerate with energy $E=4$. Primary operators with two excitations exist  for length $L \geq 6$. In Tab. \ref{tab:Z2OperatorsX} the shortest possible primary operators are listed, as well as their eigenstates and rapidities. 

We delegate a complete list of twisted-sector $\amsmathbb{Z}_2$-orbifold states with two (equal) excitations and length $L\leq 10$ to the Appendix \ref{app:Collection}.

\begin{table}
\centering
\begin{tabular}{c|c|c|c|c}
$L$ & Eigenstate & $E$ & $u_1$ & $u_2$\\
\hline
$l\in 2\amsmathbb{N}$ & $  \mathcal{B}^{1,l}_{\circ} = \mathcal{O}^{1,l}_0$ & 2 & $\pm \frac{i}{2}$ & $\mp \frac{i}{2}$ \\
$8$ & $\sqrt{6}  \mathcal{B}^{1,8}_{\pm} = \mathcal{O}^{1,8}_1 \pm i \sqrt{3}\mathcal{O}^{1,8}_2 - \mathcal{O}^{1,8}_3$ & 4 & $\pm \frac{\sqrt{3}}{2}$ & $\pm \frac{1}{2\sqrt{3}}$ \\
$10$ & $2  \mathcal{B}^{1,10}_{\pm} = \mathcal{O}^{1,10}_1 \pm i \sqrt{2}\mathcal{O}^{1,10}_2 - \mathcal{O}^{1,10}_3$ & 4 & $\frac{1}{2} \pm \frac{1}{\sqrt{2}}$ &$ -\frac{1}{2} \pm \frac{1}{\sqrt{2}}$  
\end{tabular}
\caption{Single-trace operators in the $\SU(2)_R$ and $\SU(2)_L$ sectors of the $\amsmathbb{Z}_2$-orbifold theory with lengths $L=8,10$ and two excitations. The Bethe eigenstates are given by linear combinations of the basis elements $\mathcal{O}^{k,L}_j$ defined in \eqref{eq:OpBasis}. The first operator $\mathcal{B}^{1,l}_{\circ}$ corresponds to a bound state. The spectrum of the other operators is highly degenerate with energy $E=4$ as the rapidities are related through $u_2=\frac{1}{4u_1}$.} \label{tab:Z2Operators}
\end{table}

\begin{table}
\centering
\begin{tabular}{c|c|c|c|c}
$L$ & Eigenstate & $E$ & $u_1$ & $u_2$\\
\hline
$l\in 2\amsmathbb{N}$ & $  \mathcal{C}^{1,l}_{\circ} = \mathcal{O}^{1,l}_0$ & 2 & $\pm \frac{i}{2}$ & $\mp \frac{i}{2}$\\
$6$ & $2 \mathcal{C}^{1,6}_{\pm} = \sqrt{2} \mathcal{O}^{1,6}_1 \pm i \mathcal{O}^{1,6}_2$ & 4 & $\frac{1}{2} \pm \frac{1}{\sqrt{2}}$ & $-\frac{1}{2} \pm \frac{1}{\sqrt{2}}$ \\
$8$ & $\sqrt{2} \mathcal{C}^{1,8}_{\pm} = \mathcal{O}^{1,10}_1 \pm i \mathcal{O}^{1,10}_2$ & 4 & $\pm 1 + \frac{\sqrt{3}}{2}$ &$ \pm 1 - \frac{\sqrt{3}}{2}$  
\end{tabular}
\caption{Single-trace operators in the broken $\SU(2)$ sector of the $\amsmathbb{Z}_2$-orbifold theory with lengths $L=6,8$ and two excitations. The Bethe eigenstates are given by linear combinations of the basis elements $\mathcal{O}^{k,L}_j$ defined in \eqref{eq:OpBasis}. Also in this case there exists a bound state operator $\mathcal{C}^{1,l}_{\circ}$ while the spectrum of the other operators is highly degenerate.} \label{tab:Z2OperatorsX}
\end{table}

\section{Three approaches to three-point functions} \label{Sec:ThreeApproaches}

The aim of this article is the calculation of three-point functions involving non-BPS operators in orbifold theories at weak coupling, where we can compare integrability based approaches directly to gauge theory. 

To this end, it will be necessary to consider operators at different points in 4d spacetime. Superconformal symmetry allows us to transform any configuration of three points to a collinear configuration, the only necessary transformation now being a translation along the line. Take for example the translation operator 
\begin{equation}
T=-i\epsilon_{\dot\alpha\alpha}P^{\dot\alpha\alpha}\,.
\end{equation}
Its non-trivial action on $\bar{Q}^\alpha_b$ and $S^{\dot{a}}_{\dot{\beta}}$ suggests that this translation breaks the supersymmetry remaining after light-cone gauge completely \eqref{eq:matrixrep}. However, one may salvage half the supersymmetry by acting simultaneously with an $R$-symmetry transformation, resulting in the ``twisted translation" \cite{Basso:2015zoa}
\begin{equation}\label{eq:Translation}
    \mathcal{T}=-i\epsilon_{\dot\alpha\alpha}P^{\dot\alpha\alpha}+\epsilon_{\dot a a}R^{\dot a a}\,.
\end{equation}
Since the adjective ``twisted" has already been used to describe sectors of orbifold theories, we will henceforth refer to this operation simply as \emph{translation}, keeping the additional $R$-symmetry rotation implicit. We note that a diagonal combination of the two $\PSU(2|2)$- factors remains unbroken under translation by $\mathcal{T}$. This subgroup is spanned by generators of the form
\begin{equation}
    \begin{aligned}
        \mathbf{\mathcal{R}}^{a}_{~b} &= R^{a}_{~b} + R^{\dot{a}}_{~\dot{b}}\,, \qquad &&\mathbf{\mathcal{L}}^{\alpha}_{~\beta} = L^{\alpha}_{~\beta} + L^{\dot{\alpha}}_{~\dot{\beta}}\,, \\
        \mathbf{\mathcal{Q}}_{\alpha}^{~a} &= Q_{\alpha}^{~a} + i  \epsilon_{\alpha \dot{\beta}} \epsilon^{a \dot{b}} S_{\dot{b}}^{~\dot{\beta}}\,, \qquad
        &&\mathbf{\mathcal{S}}_{a}^{~\alpha} = S_{a}^{~\alpha} + i \epsilon_{a \dot{b}} \epsilon^{\alpha \dot{\beta}} Q_{\dot{\beta}}^{~\dot{b}}\,,
    \end{aligned} \label{eq:DiagonalCharges}
\end{equation}
and is called the hexagon subalgebra in reference to the hexagon formalism \cite{Basso:2015eqa} which uses this remaining symmetry algebra to bootstrap three-point functions. We will review the hexagon formalism in Section \ref{ssec:hexagon}.

As we are restricting our attention to scalar operators, it will be useful to illustrate the effect of the translation $\mathcal{T}$ on the scalar fields explicitly. The adjoint scalar fields of $\mathcal{N}=4$ furnish an antisymmetric representation of the $\SU(4)$ $R$-symmetry, which sits in the lower-right quadrant of \eqref{eq:matrixrep}. We choose a representation 
\begin{equation}\label{eq:scalars}
    \Phi^{a\dot{a}}=\begin{pmatrix}
        0 & \bar{\Phi}_V & \bar{\Phi}_T & \Phi_L\\
        -\bar{\Phi}_V &0& \bar{\Phi}_L & \Phi_T\\
        -\bar{\Phi}_T &-\bar{\Phi}_L&0&\Phi_V\\
        - \Phi_L & -\Phi_T& -\Phi_V&0
    \end{pmatrix}\,.
\end{equation}
We then observe that the translation $\mathcal{T}$ acts non-trivially only on the following fields
\begin{equation}
    \mathcal{T}: \qquad \Phi_L\to \bar{\Phi}_V\,,\qquad \bar{\Phi}_L\to -\bar{\Phi}_V\,,\qquad \Phi_V\to\Phi_L -\bar{\Phi}_L\,.
\end{equation}
This action justifies our nomenclature in terms of the vacuum field $\Phi_V$, the longitudinal field $\Phi_L$ and the transversal field $\Phi_T$ w.r.t. the translation at hand. In $\mathcal{N}=4$ SYM we could of course make an arbitrary identification of these fields as (combinations of) the usual complex fields $X$, $Y$ and $Z$ and their conjugates. However, as we already observed in Section \ref{ssec:gauge}, the choices of vacua in the orbifold theories are distinct, so we will for now stick to this generic naming scheme.

Having outlined the necessary translation operation, we may exponentiate it to move operators to finite separations
\begin{equation}\label{eq:finite}
	\tilde{\mathcal{O}}(t) = e^{t \mathcal{T}} \mathcal{O}(0) e^{-t \mathcal{T}} \,.
\end{equation} 
Again the action on scalar fields will be of key interest.
Due to the nilpotency of $\epsilon_{a \dot{a}} R^{a \dot{a}}$ the series expansion terminates at second order
\begin{equation}
	\tilde{\Phi}^{b \dot{b}}(t) = e^{t \epsilon_{a \dot{a}} R^{a \dot{a}}} \Phi^{b \dot{b}} e^{-t \epsilon_{c \dot{c}} R^{c \dot{c}}} = \Phi^{b \dot{b}} + t \, \epsilon_{a \dot{a}} [R^{a \dot{a}},\Phi^{b \dot{b}} ] + \frac{t^2}{2} \epsilon_{a \dot{a}}  \epsilon_{c \dot{c}} [R^{c \dot{c}},[R^{a \dot{a}},\Phi^{b \dot{b}} ]] \,.
\end{equation}
In particular, we can work out the translated bosonic fields $\tilde{\Phi}^{b \dot{b}}(t)$, which read
\begin{equation}\label{eq:translatedfields}
\begin{aligned}
	\tilde{\Phi}_T(t) &= \Phi_T \,, \qquad &&\tilde{\bar{\Phi}}_T(t) = \bar{\Phi}_T \,, \\
	\tilde{\Phi}_L(t) &= \Phi_L + t\, \bar{\Phi}_V \,, \qquad &&\tilde{\bar{\Phi}}_L(t) = \bar{\Phi}_L - t \, \bar{\Phi}_V \,, \\
	\tilde{\Phi}_V(t) &= \Phi_V + t\, (\Phi_L-\bar{\Phi}_L) + t^2 \, \bar{\Phi}_V \,, \qquad &&\tilde{\bar{\Phi}}_V(t) = \bar{\Phi}_V \,.
\end{aligned}
\end{equation}
This allows us to set up a useful basis of three-point functions preserving a $\PSU(2|2)$-symmetry. Note that the translation mixes single-trace operators with certain ``descendant" states in the same superconformal multiplet, related by $R$-symmetry. This approach is tailored towards a description of states in terms of their Bethe-roots where the actual field content of the operators is secondary. Essentially one may always change the field content by adding excitations with vanishing energy and momentum to the spin-chain picture. Moving on to three-point functions, the translation $\mathcal{T}$ then ensures that an appropriately $R$-rotated descendant is present at each point to make the overall configuration an $R$-symmetry singlet \cite{Basso:2015zoa}. We can then identify the structure constants for all $R$-symmetry conserving combinations.

We will now use three approaches to evaluate them at tree level: field theory, spin-chain overlaps and the hexagon formalism. We will review these methods and their application to $\mathcal{N}=4$ SYM in the following, commenting on their applicability to orbifold theories.

\subsection{Wick contractions}
In the field-theory calculation we simply perform planar tree-level Wick contractions.
As observed above, the translation \eqref{eq:Translation} rotates some of the fields into each other. We may determine the effective propagators of the translated fields \eqref{eq:translatedfields}
by performing standard Wick contractions:
\begin{equation}\begin{split}
	\braket{\tilde{\Phi}_T(t_i) \tilde{\bar{\Phi}}_T(t_j)} &= \frac{1}{(t_i-t_j)^2} \,, \quad \braket{\tilde{\Phi}_L(t_i) \tilde{\bar{\Phi}}_L(t_j)} = \frac{1}{(t_i-t_j)^2} \,, \\ \braket{\tilde{\Phi}_L(t_i) \tilde{\Phi}_V(t_j)} &= \frac{1}{t_i-t_j} \,, \quad \braket{\tilde{\bar{\Phi}}_L(t_i) \tilde{\Phi}_V(t_j)} = -\frac{1}{t_i-t_j} \,, \quad \braket{\tilde{\Phi}_V(t_i) \tilde{\Phi}_V(t_j)} = 1\,. \label{eq:Propagators}
\end{split}\end{equation}
A tree-level evaluation of correlation functions can now be performed by doing Wick contractions and using the $\SU(N)$ trace rules 
\begin{equation}
\begin{aligned}
    \tr \, (T^a A) \, \tr \, (T_a B) &= \left(\tr \, (AB) - \frac{1}{N} \tr \, (A) \tr \, (B) \right)\,, \\
    \tr \, (T^a A \,T_a B) &= \left( \tr \, (A) \tr \, (B) - \frac{1}{N} \tr \, (AB) \right)\,, \\
\end{aligned}
\end{equation}
where $T^a$ are the generators of $\SU(N)$. This method is tailored towards $\mathcal{N}=4$ SYM and moving on to orbifold theories, we will have to take into account the more intricate gauge structure. One may decompose the $\SU(MN)$ representations as in \eqref{eq:GaugeMatrix} and contract strictly fields which transform under the same gauge groups. This yields the same results as a more implicit prescription due to \cite{Ideguchi:2004wm} where we work at the level of $\SU(MN)$-representations and keep track of commutations with the twist operator $\gamma$.

\subsection{Spin-chain overlap} \label{Sec:SpinChainOverlap}
The spin-chain overlap method makes use of the coordinate Bethe Ansatz for the evaluation of correlation functions.
In \cite{Escobedo:2010xs} an elegant integrability-based framework was developed for three-point functions.
The evaluation can even be used to read off the tree-level hexagon form factor, \cf the construction in \cite{Basso:2015zoa} for one-magnon states. In principle it can also be lifted to loop corrections by inserting Hamiltonian densities at the splitting points of the spin chains \cite{Okuyama:2004bd,Alday:2005nd}. 

For the purposes of this paper, if suffices to use blunt tools. To set the scene let us consider correlators featuring two non-BPS operators and one BPS operator in $\mathcal{N}=4$ SYM. The BPS operator we refer to as \emph{reservoir} and it consists of the field $\tilde{\Phi}_V$ only. Hence, it can have contractions with vacuum fields and longitudinal excitations from the non-BPS operators.

The non-BPS operators are described by their wave function \eqref{eq:BetheState}. For example the two-particle state may be abbreviated as 
\begin{equation}
    \psi^{p,q}_{n,m} \equiv  e^{i(n p + m q)} + e^{i(n q + m p)} S_{p,q} \,.
\end{equation}
The overlap can be calculated by summing the product of the two operator wave functions over all possible contractions. Furthermore, we will dress the phase factors with their propagators \eqref{eq:Propagators}, \eg~when we contract an excitation $\tilde{\Phi}_L$ sitting at point $t_1$ on a vacuum field $\tilde{\Phi}_V$ from an operator at $t_2$, the corresponding factor is $\tfrac{1}{t_{1}-t_2}$. Since the translation \eqref{eq:Translation} mixes vacuum and longitudinal fields, we need to distinguish different cases in the evaluation. 

\paragraph{Transversal excitations.}
\begin{figure}
    \centering
    \includegraphics[width = 1\textwidth]{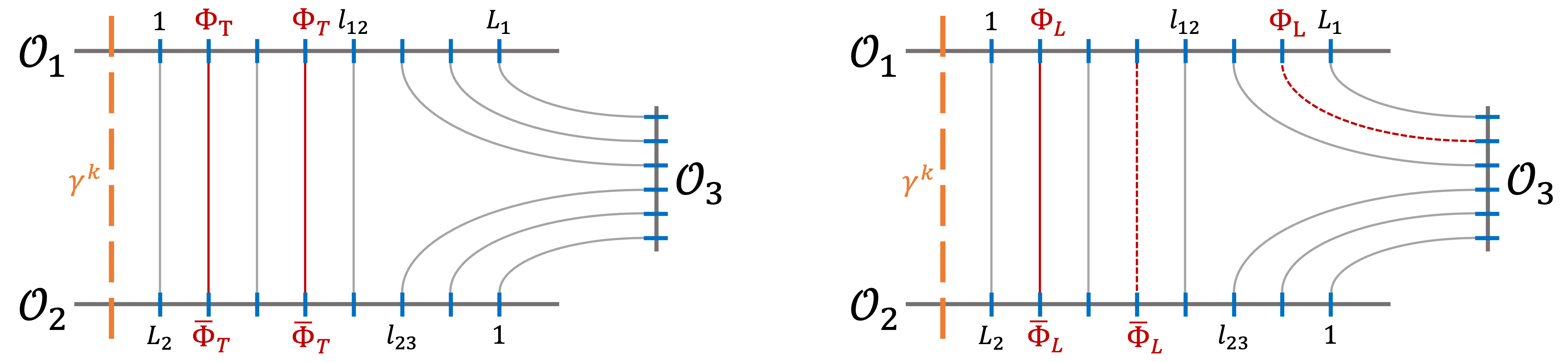}
    \caption{Three-point functions can be calculated from spin-chain overlaps summing over all possible contractions. Left: For the transversal excitations only conjugate fields can be contracted. Right: The longitudinal excitations can either be contracted on the translated vacuum or on their conjugate counterpart (contact terms). We also marked the position of the twist operator in a twisted-twisted-untwisted three-point function in the orbifold theory.} 
    \label{Fig:ThreePointOverlap2Ex}
\end{figure}
We consider two operators carrying transversal excitations. The first operator is at $t_1$, has length $L_1$ and two excitations $\Phi_T$, while the second is at $t_2$ with length $L_2$ and two excitations $\bar{\Phi}_T$. Further, we consider a vacuum of length $L_3$ at $t_3$. 
In the spin-chain picture, the excitations can propagate freely over the chain. Since the fields are transversal, $\Phi_T$ can only be contracted on an excitation $\bar{\Phi}_T$.
Hence, in the overlap, we only have \emph{contact terms}, as depicted on the left side of Fig. \ref{Fig:ThreePointOverlap2Ex}. In the notation of \cite{Basso:2015zoa}, the overlap is then given by
\begin{equation}
    C^{\bullet \bullet \circ}_{\Phi_T,\bar{\Phi}_T} = N_1 N_2 \sum_{1 \leq n_1 < n_2 \leq \ell_{12}} \frac{\psi^{p_1,p_2}_{n_1,n_2} * \psi^{p_3,p_4}_{L_2-n_2+1,L_2-n_1+1}}{t_{12}^4}\,, \label{eq:OverlapX}
\end{equation}
where $\ell_{12}=(L_1+L_2-L_3)/2$ is the number of tree-level propagators between the operators at $t_1$ and $t_2$ and $t_{12}=t_1-t_2$ their physical distance.
The factors $N_1$ and $N_2$ are the normalisation factors of the respective operators as given in eq. \eqref{eq:BetheNorm}.

\paragraph{Longitudinal excitations.} 
As the longitudinal excitations $\tilde{\Phi}_L$ and $\tilde{\bar{\Phi}}_L$ can also be contracted on the vacuum, we have to consider two cases; either both operators carry different excitations, \ie $\tilde{\Phi}_L$ and $\tilde{\bar{\Phi}}_L$, or they carry only one kind. 
In the first case, the fields can be contracted on the translated vacuum $\tilde{\Phi}_V$ as well as producing contact terms, as depicted on the right side of Fig. \ref{Fig:ThreePointOverlap2Ex}.  Hence the excitations move independently over the spin chain and the overlap is given by
\begin{equation}
\begin{aligned}
    C^{\bullet \bullet \circ}_{\Phi_L,\bar{\Phi}_L} = N_1 N_2 &\left( \sum_{1 \leq n_1 < n_2 \leq \ell_{12}} \frac{\psi^{p_1,p_2}_{n_1,n_2}}{t_{12}^2} + \sum_{\substack{1 \leq n_1 \leq \ell_{12} \\ \ell_{12} < n_2 \leq L_1}} \frac{\psi^{p_1,p_2}_{n_1,n_2}}{t_{12}t_{13}} + \sum_{\ell_{12} < n_1 < n_2 \leq L_1} \frac{\psi^{p_1,p_2}_{n_1,n_2}}{t_{13}^2}\right) \times \\
    &\left( \sum_{1 \leq m_1 < m_2 \leq \ell_{23}} \frac{\psi^{p_3,p_4}_{m_1,m_2}}{t_{23}^2} + \sum_{\substack{1 \leq m_1 \leq \ell_{23} \\ \ell_{23} < m_2 \leq L_2}} \frac{\psi^{p_3,p_4}_{m_1,m_2}}{t_{23}t_{21}} + \sum_{\ell_{23} < m_1 < m_2 \leq L_2} \frac{\psi^{p_3,p_4}_{m_1,m_2}}{t_{21}^2}\right)\,. \label{eq:OverlapYYB}
\end{aligned}
\end{equation}

Finally, we consider the case with excitations $\tilde{\Phi}_L$ on both operators $C^{\bullet \bullet \circ}_{\Phi_L,\Phi_L}$. Since there is no propagator between two fields $\tilde{\Phi}_L$, there cannot be any contact terms. Therefore, we have to subtract these terms. For the reader's convenience we give this rather bulky expression explicitly in Appendix \ref{App:ContactTerms} .

If we want to apply this technology to orbifold theories, we are faced with the same subtleties about gauge representations we had to consider in the approach using Wick contractions. Furthermore, when we consider three-point functions involving twisted operators, we have to take into account the commutation of fields with the twist operator. Take for example the three-point function of two non-BPS twisted operators (with twist numbers $k$ and $M-k$, respectively) and one BPS untwisted operator. In each twisted operator, the twisted boundary condition is placed between the sites $L$ and $1$, \ie depending on the sector a factor $\omega$ is picked up, when an excitation moves from site $L$ to $L+1$ (\cf Fig. \ref{Fig:ThreePointOverlap2Ex}).
It seems worth noting that with this placement no explicit twist appears in the following evaluation. The twist only enters through the Bethe roots as solutions of the twisted Bethe equations. We will see this pattern reemerge in discussing the hexagon formalism in the next Section.

\subsection{The hexagon formalism}\label{ssec:hexagon}
Although the hexagon formalism was already outlined in the introduction, let us provide some formulae for the evaluation of a sample three-point function in $\mathcal{N}=4$ SYM. For simplicity, we consider the correlator of one non-BPS operator $\mathcal{B}$ with two BPS operators $\mathcal{O}$. 
The operators are described in the integrability picture, so cutting an operator corresponds to cutting a Bethe state. Considering for instance two excitations, there are four partitions $\alpha,\bar{\alpha}$ with $\alpha \cup \bar{\alpha}= \{u_1,u_2\}$ over which the hexagon has to be summed. The sum is weighted by the splitting factor \cite{Escobedo:2010xs,Basso:2015zoa}
\begin{equation}
    \rho_{\ell}(\alpha,\bar{\alpha}) = \prod_{j \in \bar{\alpha}} e^{i p_j \ell} \prod_{\substack{j,k \\ k \in \alpha}} S_{p_j,p_k}\,. \label{eq:UndefWeightFactor}
\end{equation}
The parameter $\ell$ measures the length of the sub-chains after cutting. 
Moreover, for the three-point function we have
\begin{equation}
\braket{\mathcal{B}_1 \mathcal{O}_{L_2} \mathcal{O}_{L_3} } 
= \sqrt{ \frac{L_1 L_2 L_3}{\mathcal{G} S_{1,2}} } 
\sum_{\alpha \cup \bar{\alpha} = \{u_1, u_2\}} 
\frac{(-1)^{|\bar{\alpha}|} t_{23}}{t_{12}t_{13}} \, \rho_{\ell_{12}}(\alpha, \bar{\alpha}) 
\, \braket{ \h | \alpha} \, \braket{ \h | \bar{\alpha} } \,, \label{eq:BKVHexagon}
\end{equation}
where the hexagon form factor is $\braket{ \h | \alpha}$. Its explicit evaluation reveals a combination of the matrix elements of the $\su(2|2)$ S matrix \cite{Beisert:2005tm} and the scalar hexagon dressing factor \cite{Basso:2015zoa}. For instance, at tree-level the zero-, one- and two-particle form factors for longitudinal excitations are given by 
\begin{equation}
    \braket{\h | \{\}}=1\,,  \quad \braket{\h | \{\Phi_L(u)\}}=1\,, \quad \braket{\h | \{\Phi_L(u_1) ,\Phi_L(u_2) \}}=\frac{u_1 -u_2}{u_1-u_2-i}\,.
\end{equation}
For the evaluation of more general form factors we refer to the original literature \cite{Basso:2015zoa}\footnote{A further review is given in \cite{lePlat:2023fca}, of which we use, in particular, the conventions fixed in Appendix A.}.
The edge width $\ell_{12}$ is given by the number of tree-level propagators between the operators $\mathcal{B}$ and $\mathcal{O}_{L_2}$. As before it is given by $\ell_{12} = (L_1 + L_2 - L_3)/2$. 
The expression in \eqref{eq:BKVHexagon} is valid for asymptotically large operators. To compute finite-size corrections, full sets of virtual particles need to be inserted on the edges of the cut worldsheet \cite{Basso:2015zoa}. This Lüscher-like approach allows to evaluate the finite-size contributions order by order in the coupling.

The generalisation to more operators carrying excitations is straightforward, though one has to deal with a growing number of partitions. To evaluate the form factor, all the excitations need to be brought to the same edge. This is achieved through crossing transformations. Working out the hexagon formalism for orbifold theories will be the focus of the next Section. The emerging formalism is very similar to the one of $\mathcal{N}=4$, with only two differences: first of all, the rapidities entering know about the orbifolding as they are solutions to the twisted Bethe equations. Secondly, when cutting the Bethe state one has to keep track of the twist. Depending on the considered sector and its Bethe equations, excitations may pick up twist factors when they move from one hexagon to another. 

Let us illustrate this effect with a simple two-magnon state and then generalise to arbitrary states. In our conventions (see Fig. \ref{Fig:ThreePointOverlap2Ex} and \eqref{eq:OpBasis}) the twist $\gamma^k$ is located between the $L$-th and the first site of the chain. Cutting such a state into chain segments of length $\ell$ and $\bar{\ell}=L-\ell$, the excitations move from one chain to the other without picking up extra twist factors $\omega$. Explicitly and similarly to \eqref{eq:UndefWeightFactor} for a two-particle state we have
\begin{equation}
\begin{aligned}
    \mathcal{B} \,\, \sim \,\, &\ket{p_1,p_2}_{\ell} \otimes \ket{0}_{\bar{\ell}} + e^{i p_2 \ell} \ket{p_1}_{\ell} \otimes \ket{p_2}_{\bar{\ell}} \\
    &\quad + e^{i p_1 \ell} S_{p_1,p_2} \ket{p_2}_{\ell} \otimes \ket{p_1}_{\bar{\ell}} + e^{i (p_1+p_2) \ell} \ket{0}_{\ell} \otimes \ket{p_1,p_2}_{\bar{\ell}} \,.
\end{aligned}
\end{equation}
Here, the excitations are moved ``to the right" along the chain by $\ell$ sites from the first to the second segment.
We can now multiply this equation by $e^{-i(p_1+p_2)L} \omega^{kpL}=1$ and use the Bethe equations \eqref{eq:BAE} as well as the orbifold invariance condition $\omega^{k (L-2)p+2k q}=1$ to obtain
\begin{equation}
\begin{aligned}
    \mathcal{B} \,\, \sim \,\, &\ket{p_1,p_2}_{\ell} \otimes \ket{0}_{\bar{\ell}} +  \omega^{k(p-q)} S_{p_1,p_2} e^{-i p_2 \bar{\ell}} \ket{p_1}_{\ell} \otimes \ket{p_2}_{\bar{\ell}} \\
    &\quad +\omega^{k(p-q)} e^{-i p_1 \bar{\ell}} \ket{p_2}_{\ell} \otimes \ket{p_1}_{\bar{\ell}} + \omega^{2k(p-q)} e^{-i (p_1+p_2) \bar{\ell}} \ket{0}_{\ell} \otimes \ket{p_1,p_2}_{\bar{\ell}}\,.
\end{aligned}
\end{equation}
This can be interpreted as the excitations being moved to the left by $\bar{\ell}$ sites. When moving over the twist $\gamma^k$ between the $L$-th and the first site, each magnon picks up their respective twist factor $\omega^{k(p-q)}$. We thus observe that internal consistency of the Bethe equations immediately dictates the appropriate twist factors when moving magnons across the hexagons. In general, shifting magnons either to the left or to the right results in splitting factors which are related as
\begin{equation}
    \prod_{j \in \bar{\alpha}} e^{i p_j \ell} \prod_{\substack{j<k \\ k \in \alpha}} S_{p_j,p_k} = \omega^{k |\bar{\alpha}|(p-q)}\prod_{j \in \bar{\alpha}} e^{-i p_j \bar{\ell}} \prod_{\substack{j>k \\ k \in \alpha}} S_{p_k,p_j}\,,
\end{equation}
where the additional twist factor is apparent. In an explicit calculation we may decide on a direction to shift the magnons and then evaluate whether twist factors are needed according to these consistency conditions. In the following sections, we will always assume movement of magnons to the right.

Although the overall accounting of the twist factors is fixed by consistency, the basis choice \eqref{eq:OpBasis} and position of the twist operator is conventional (at least at tree-level). Different conventions for the placement of twist operators within the traces are related to ours by overall phase factors which cancel out in  appropriately normalised three-point functions. One may even consider more exotic conventions, such as splitting the overall twist operator as $\gamma^k=\gamma^{k_0}\gamma^{k_\ell}$ and keeping the twist $\gamma^{k_0}$ between sites $L$ and one, while we move $\gamma^{k_\ell}$ by $\ell$ sites, \ie to the right of the first segment. This results in a phase $\omega^{-k_\ell (p \ell +(q-p)(|\alpha|+|\bar\alpha|))}$. Additionally, the magnons now pick up a twist factor $\omega^{k_\ell |\bar{\alpha}|(q-p)}$ when moving over the edge with twist $\gamma^{k_{\ell}}$. This allows us to formulate a splitting factor for general configurations of twist lines $\gamma^{k_0}$ and $\gamma^{k_\ell}$ between the two segments\footnote{Note that this equation also holds if no twist operator is moved to the right, \ie $k_{\ell}=0$.}
\begin{equation}
\begin{aligned}
    \rho^{k_{\ell}}_{\ell}(\alpha,\bar{\alpha}) &= \omega^{-k_\ell (p \ell +(q-p)|\alpha|)} \omega^{k_\ell |\bar{\alpha}|(q-p)} \prod_{j \in \bar{\alpha}} e^{i p_j \ell} \prod_{\substack{j<k \\ k \in \alpha}} S_{p_j,p_k} \\
    &= \omega^{-k_\ell (p \ell +|\alpha| (q-p))} \prod_{j \in \bar{\alpha}} e^{i p_j \ell} \prod_{\substack{j<k \\ k \in \alpha}} S_{p_j,p_k}\,. \label{eq:OrbiWeightFactor}
\end{aligned}
\end{equation}
Moving the magnons to the left instead, respective factors for crossing the twist $\gamma^{k_0}$ are picked up. Again, the Bethe equations together with the orbifold invariance ensure the equivalence when moving the magnons.

The above self-consistent prescription for orbifolded hexagons at tree-level is directly motivated and supported by the twisted Bethe equations known from the spectral problem \cite{Ideguchi:2004wm,Beisert:2005he}\,. It is therefore straight-forward to extend our prescription to higher-rank sectors and other twisted scenarios like the $\beta$- and $\gamma$-deformations of $\mathcal{N}=4$ SYM \cite{Eden:2022ipm}. The same strategy for computing the three-point functions may be applied: We  choose a basis of operators solving the twisted Bethe equations, determine where the twist lines are positioned and whether twist factors analogous to \eqref{eq:OrbiWeightFactor} have to be included in the splitting factor, then apply the usual hexagon construction. In the following we will consider a few simple $\SU(2)$ sectors for simplicity but would like to emphasise that our prescription can be generalised directly. We will comment on this possibility again in the conclusion.

\section{The hexagon for orbifolds} \label{Sec:Z2Hexagon}

In this Section, we will set up and evaluate three-point functions in orbifold theories via the hexagon formalism \cite{Basso:2015zoa}. We will use the simplest $\mathcal{N}=2$ $\amsmathbb{Z}_2$-orbifold theory as an instructive example but comment on the generalisation to $\amsmathbb{Z}_M$ orbifolds. In order to check our computation, we compare the results explicitly to tree-level gauge theory. This comparison is slightly complicated by the fact that the translation  $\mathcal{T}$ \eqref{eq:Translation} which preserves a $\PSU(2|2)$ symmetry in $\mathcal{N}=4$ SYM and is used to set up three-point function calculations does not commute with the orbifold twist. 
In fact we can see by comparing with \eqref{eq:adjoint} and \eqref{eq:bifundamental} that all supersymmetry is broken when choosing the adjoint gauge ($\Phi_V=Z$, $\Phi_L=Y$) and only a diagonal $\PSU(1|1)$ group remains when choosing the bifundamental gauge ($\Phi_V=X$, $\Phi_L=Y$).\footnote{We have been careful to separate the notion of vacuum, transversal and longitudinal fields $(\Phi_V,\Phi_T,\Phi_L)$ from  explicit fields $(X,Y,Z)$, which transform differently under the orbifold action \eqref{eq:orbifold action}. The ``gauge choices" performed here determine the relative orientation of translation and orbifold action and lead to physically distinct scenarios. }

When applying a naive translation \eqref{eq:finite} to a generic state, we therefore generate a bunch of states that do not close into traces of the same gauge group and have to be discarded as unphysical.\footnote{In the language of \cite{Bertle:2024djm}, the finite translation is uplifted to a finite groupoid transformation.} Alongside these we also generate physical states that do not belong to the same superconformal multiplet anymore. This results in a superposition of three-point functions involving various multiplets. The hexagon formalism now seems to yield the structure constants of individual components of this superposition, which are $R$-symmetry singlets under the reduced $R$-symmetry. However, a comparison to the gauge theory is now complicated by the plethora of multiplets involved. 

If we want to test the hexagon proposal against gauge theory and spin-chain overlaps, we may now follow two possible strategies. We can keep setting up three-point functions by employing the translation $\mathcal{T}$ and consider operators for which the descendant structure is reasonably simple. This approach shows great benefit in the bifundamental gauge where a $\PSU(1|1)$ is preserved. The other possibility is to abandon the translation operator $\mathcal{T}$ and to ensure $R$-symmetry conservation ``by hand". To this end we pick particular operators inserted at $0$, $1$ and $\infty$ along a line in spacetime and make sure they can be contracted completely. This comes at the benefit of probing individual structure constants directly without the need to disentangle a superposition of three-point functions, but requires a large amount of bookkeeping when tabulating three-point functions. This approach is natural to the discussion of the adjoint vacuum where all supersymmetry is broken anyway, so both strategies would require the same amount of effort.

We will first consider the structure constants of $\SU(2)$ sectors in the $\mathcal{N}=2$ $\amsmathbb{Z}_2$-orbifold theory in bifundamental gauge. In this special case, we may exploit the additional $\SU(2)$-symmetry between the $X$ and $Y$ fields. This makes a discussion following the first strategy feasible. If we instead choose the adjoint gauge, we face a less symmetric situation in which the second strategy seems more straightforward. We then comment on $\amsmathbb{Z}_M$ orbifolds and show how particularly easy three-point functions can be evaluated and matched. In all explicit examples, we find complete agreement of Wick contractions, spin-chain overlap and the hexagon formalism.

\subsection{$\amsmathbb{Z}_2$ orbifolds in bifundamental gauge}\label{ssec:Simplesec}

The operator spectrum of the $\amsmathbb{Z}_2$ orbifold consists of an untwisted and a single twisted sector. This implies that the superselection rule \eqref{eq:superselection} only allows for two possible sector combinations in the three-point function
\begin{equation}
    \braket{\mathcal{O}^0\mathcal{O}^0\mathcal{O}^0}\qquad \text{and}  \qquad\braket{\mathcal{O}^1\mathcal{O}^1\mathcal{O}^0}.
\end{equation}
We will be particularly interested in the second case involving two twisted and one untwisted operator.

\begin{figure}[h]
    \centering
    \includegraphics[width = .8\textwidth]{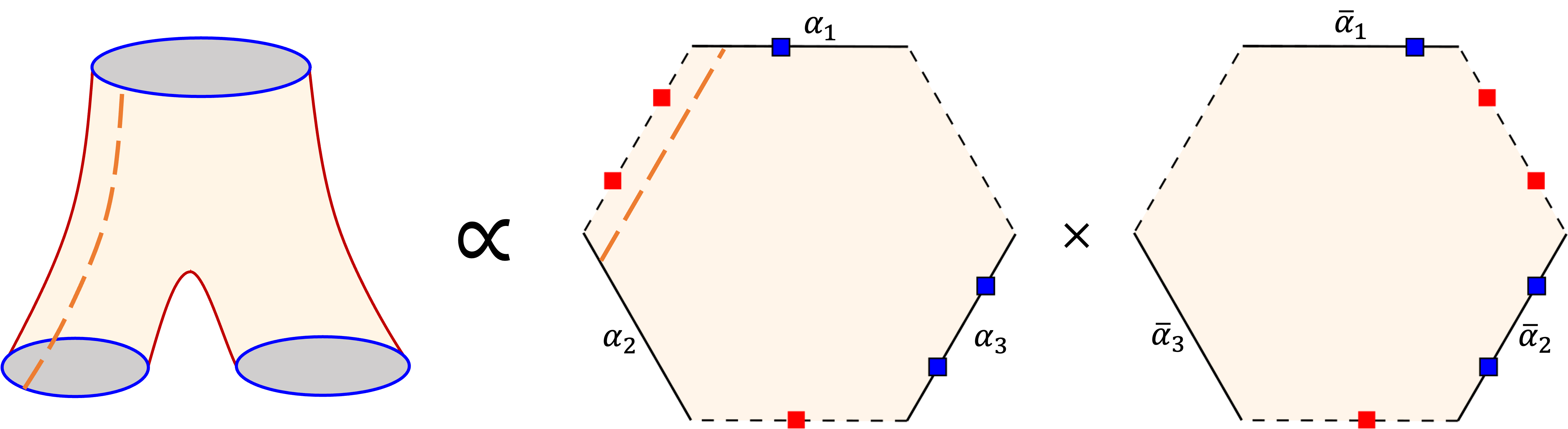}
    \caption{For three-point functions in orbifold theories involving twisted operators, we have to introduce twist operators into the external traces. Here we consider a twisted-twisted-untwisted correlator and extend the twist along the orange line.
    Moving the twist over the spin chain, it can be moved to either hexagon. Magnons may pick up a twist factor $\omega^k$ when they travel over an edge carrying twist. In scenarios with three twisted operators, the twist lines meet as in Fig. \ref{Fig:Twistvertex} and generate the superselection rule \eqref{eq:superselection}. One may then resolve the vertex into two separate twist lines connecting e.g. $\alpha_1$ with $\alpha_2$ and $\alpha_1$ with $\alpha_3$, respectively.}
    \label{Fig:TwistHexagon}
\end{figure}

\begin{figure}[h]
    \centering
    \includegraphics[width = .45\textwidth]{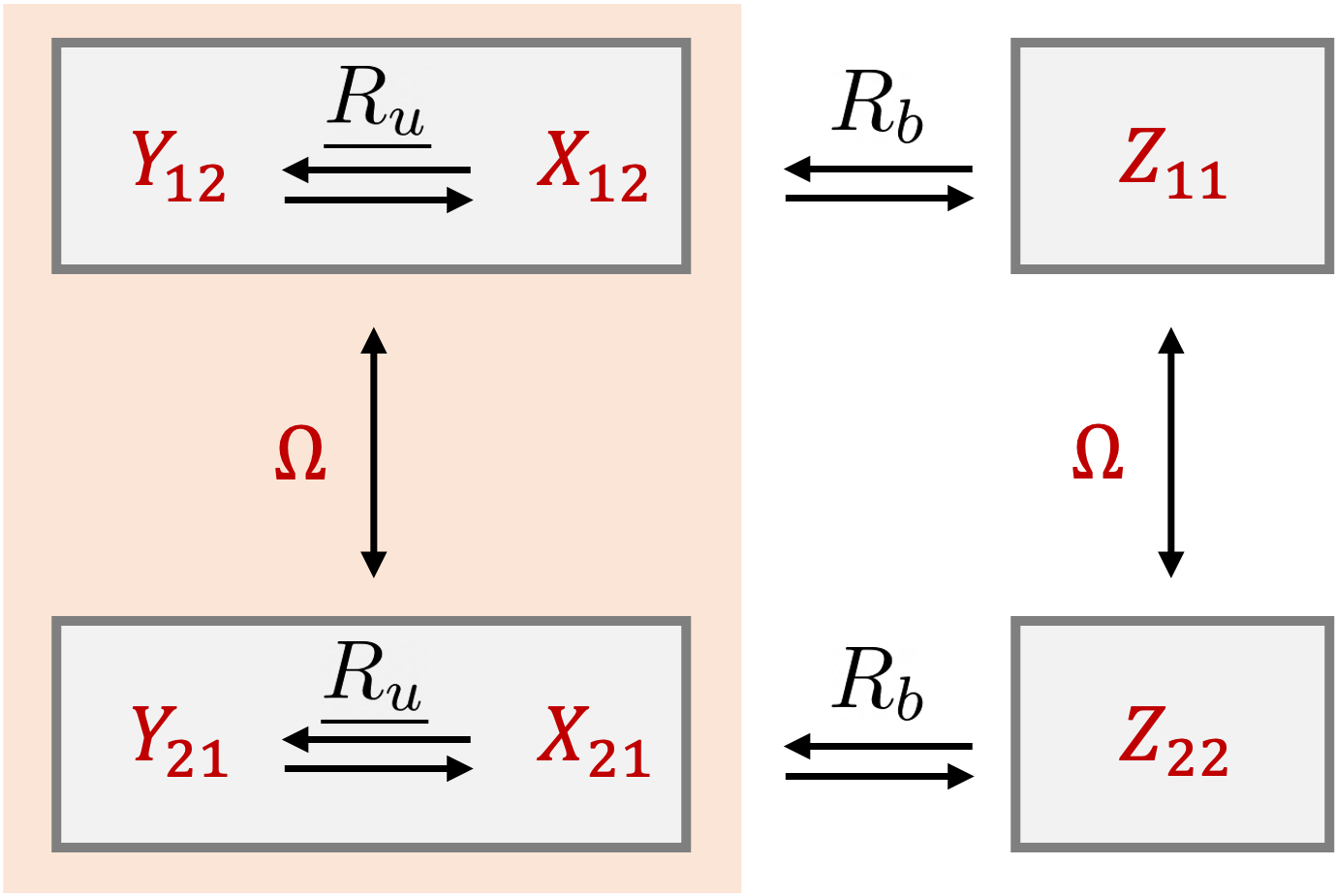}
    \caption{The algebroid structure of the bosonic part of the hexagon subalgebra \eqref{eq:DiagonalCharges} (orange) for the $\amsmathbb{Z}_2$ orbifold. Broken and unbroken $R$-symmetry generators $R_b$ and $\underline{R_u}$ are taken from \eqref{eq:gens}. }
    \label{Fig:Algebroid}
\end{figure}

Choosing the bifundamental gauge, we have the following sets of unbroken and broken $R$-symmetry generators (\cf \eqref{eq:bifundamental})
\begin{equation}\label{eq:gens}
\begin{aligned}
    &\text{unbroken:} \qquad &&\{R^{1}{}_1,R^{2}{}_2,R^{{\dot{1}}}{}_{\dot{1}},R^{{\dot{2}}}{}_{\dot{2}} ,\underline{R^{1}{}_{\dot{1}},R^{{\dot{1}}}{}_1,R^{2}{}_{\dot{2}},R^{{\dot{2}}}{}_2}\} \,, \\
    &\text{broken:} \qquad &&\{R^{1}{}_2,R^{2}{}_1,R^{{\dot{1}}}{}_{\dot{2}},R^{{\dot{2}}}{}_{\dot{1}},R^{2}{}_{\dot{1}},R^{{\dot{1}}}{}_2,R^{1}{}_{\dot{2}},R^{{\dot{2}}}{}_1\} \,,\\
\end{aligned}
\end{equation}
where the underlined generators are only unbroken in the $\amsmathbb{Z}_2$ orbifold due to the symmetry enhancement $\omega=\omega^{-1}=-1$ in \eqref{eq:bifundamental}.
As the $R$-symmetry component of the translation $\mathcal{T}$ \eqref{eq:Translation} is build from precisely these underlined generators, it preserves the multiplet structure of scalar operators without any mixing, see Fig. \ref{Fig:Algebroid}. It is easy to see that any other gauge choice would lead to a mixing of representations under translation.
By fixing the translation operator $\mathcal{T}$ \eqref{eq:Translation} and the vacuum field $\Phi_V=X$ we have implicitly also fixed the longitudinal field $\Phi_L=Y$. In terms of the $\SU(2N)$ matrices (analogously to the notation introduced in \eqref{eq:GaugeMatrix}), we may decompose the translated scalar fields as
\begin{equation}
\begin{aligned}
\tilde{Z}(t) &= \begin{pmatrix}
Z_{11} & 0 \\
0 & Z_{22}
\end{pmatrix}, \quad
\tilde{Y}(t) = \begin{pmatrix}
0 & Y_{12} + t \bar{X}_{12} \\
Y_{21} + t \bar{X}_{21} & 0 
\end{pmatrix}\,, \\
\tilde{X}(t) &= \begin{pmatrix}
0 & X_{12} + t (Y_{12} - \bar{Y}_{12}) + t^2 \bar{X}_{12} \\
X_{21} + t (Y_{21} - \bar{Y}_{21}) + t^2 \bar{X}_{21} & 0
\end{pmatrix}\,. \label{eq:Z2TwistCompFields}
\end{aligned}
\end{equation}
We indeed observe that there is no mixing between bifundamental and adjoint fields.
Furthermore, the translated fields satisfy the same twist relations as the untranslated fields. Explicitly, we have
\begin{equation}
	\gamma^\dagger \tilde{X} \gamma = - \tilde{X}\,, \quad \gamma^\dagger \tilde{Y} \gamma = - \tilde{Y}\,, \quad \gamma^\dagger \tilde{Z} \gamma = + \tilde{Z} \, .\label{eq:TwistedTranslatedOrbiTwist}
\end{equation}
Therefore, translating an operator from the twisted sector results in another twisted sector operator
\begin{equation}
	\mathcal{O}^1(t) = e^{t \mathcal{T}} \mathcal{O}^1(0) e^{-t \mathcal{T}} =  \tr( \gamma \, \tilde{\Phi}^{a_1 \dot{a}_1} \tilde{\Phi}^{a_2 \dot{a}_2} \dots \tilde{\Phi}^{a_L \dot{a}_L} )\,,
\end{equation}
with the fields $\tilde{\Phi}$ satisfying the same twisted boundary conditions. In this sense the bifundamental gauge is well-behaved because translated operators transform homogeneously under the orbifold projection, meaning that no unphysical configurations need to be projected out. 
We may compute some simple three-point-functions by inserting $\SU(2)$-sector operators at the origin and translating them to the points of interest. As they will remain in their multiplet under the action of the translation, performing gauge-theory calculations for correlators involving such operators is straightforward.

The hexagon form factor can be bootstrapped from symmetry \cite{Basso:2015zoa}. The symmetry preserved by $\mathcal{T}$ is a diagonal $\psu(2|2)_D$ subalgebra -- the hexagon subalgebra introduced in \eqref{eq:DiagonalCharges}. The $R$-symmetry part of the hexagon subalgebra is given by
\begin{equation}
    \{R^{1}{}_1+ R^{\dot{1}}{}_{\dot{1}},R^{2}{}_2+ R^{{\dot{2}}}{}_{\dot{2}},R^{1}{}_2+ R^{{\dot{1}}}{}_{\dot{2}},R^{2}{}_1+ R^{{\dot{2}}}{}_{\dot{1}} \}\,,
\end{equation}
and we see that this diagonal $\su(2)_D$ is broken -- and hence the term \emph{hexagon subalgebroid} might be more suitable \cite{Bertle:2024djm}. 

We shall now assume that the hexagon form factor is blind to the colour structure of the excitations involved. Observations indicating this were made in \cite{Eden:2017ozn}, where the hexagon result had to be dressed by $\SU(N)$ colour factors to reproduce field-theory results. 
Imposing that the hexagon form factor preserves the (otherwise broken) $\PSU(2|2)$ symmetry leads to the same bootstrap equations as in \cite{Basso:2015zoa}
\begin{equation}
    \braket{\h| g|\Psi} = 0\,, \qquad \forall \, g \, \in \, \psu(2|2)_D\,,
\end{equation}
for \emph{any} (off-shell) state $\ket{\Psi}$. Omitting the colour indices, the algebroid coproduct \eqref{eq:AlgebroidCoproduct} becomes the standard coproduct. This fixes the one- and two-particle form factors and can be generalised to arbitrary many particles. It takes the same form as in \cite{Basso:2015zoa} and can be written relatively simple in terms of the S-matrix elements \cite{Beisert:2005tm} and the hexagon dressing phase. 
In order to construct three-point functions, we need to glue together two hexagon form factors as in \eqref{eq:BKVHexagon}, which is demonstrated in Fig.  \ref{Fig:TwistHexagon}. As mentioned above, when the state is cut, we need to keep track of magnons moving through the twist operator, resulting in additional twist factors in \eqref{eq:OrbiWeightFactor}. By moving the twist over the spin chain, we can make sure that it is for instance always on the front hexagon (see also Fig. 1 in \cite{Ferrando:2025qkr}). This and the modified Bethe roots are the only changes we impose on the hexagon formalism. 

Under this premise, we can exactly reproduce tree-level gauge theory as well as spin-chain overlap predictions, as we will show momentarily. We take this as first evidence in favour of the proposal made in Section \ref{sec:Introduction}. This result is not surprising taking into account the observations from spin-chain overlaps in Section \ref{Sec:SpinChainOverlap}, where the twist also only enters implicitly through the Bethe roots.

\subsubsection{Three-point functions with two twisted operators carrying magnons}
Having set up the formalism in the preceding sections, let us present here some explicit results for three-point functions between two excited states $\mathcal{B}_{1}$, $\mathcal{B}_{2}$ and one groundstate $\mathcal{O}_L$ of length $L$. Here, we will consider $\SU(2)$-sector states introduced in Sec. \ref{sec:SU2Sectors} and App. \ref{app:Collection}.
We choose to bring the excitations over the same edge $\ell_{12}$.
The explicit hexagon formula reads 
\begin{equation}\begin{split}
&\braket{\mathcal{B}_1\mathcal{B}_{2} \mathcal{O}_{L} } 
\sim 
\delta_{0,K\, \mathrm{mod}\, M}\times\phantom{\Bigg(}\\
&\qquad\sum_{\substack{\alpha_1 \cup \bar{\alpha}_1 = \{u_1, u_2\} \\ \alpha_2 \cup \bar{\alpha}_2 = \{u_3, u_4\}}}
\frac{(-1)^{|\bar{\alpha}_1|+|\alpha_2|}}{t_{12}^2}\rho_{\ell_{12}}(\alpha_1, \bar{\alpha}_1) \, 
\rho_{\ell_{12}}(\bar{\alpha}_2, \alpha_2)
\,\braket{ \h | \alpha_1, \alpha_2, \{\} } \,
\braket{ \h | \bar{\alpha}_1, \{\}, \bar{\alpha}_2 } \,, \label{eq:Hex3ptPartition}\end{split}
\end{equation}
where the physical hexagon edges $\alpha_i,\bar{\alpha}_i$ are indicated as in Fig. \ref{Fig:TwistHexagon}
and the factor of proportionality is given by the standard normalisation $\sqrt{\frac{L_1 L_2 L}{\mathcal{G}_1 S_{1,2} \mathcal{G}_2 S_{3,4}}}$ for the hexagon. Due to the Bethe equations \eqref{eq:BetheEq} we can use the splitting factor $\rho_{\ell_{12}}(\alpha, \bar{\alpha}) $ from \eqref{eq:UndefWeightFactor}. Further, as mentioned in \eqref{eq:superselection}, the correlator has to have vanishing total twist $K\text{ mod }M = 0$ where $K=k_1+k_2+k_3$ is the sum of the twists $k_i$ of the individual operators.

\paragraph{Longitudinal excitations.}
\begin{table}[t]
\centering
\begin{tabular}{c||c|c|c|c|c}
Correlators & $L=2$ & $4$ & $6$ & $8$ & $10$\\
\hline \hline
$\braket{\mathcal{B}_{\pm}^{1,8} \mathcal{B}_{\pm}^{1,8} \mathcal{O}_L}$ & $-4 \sqrt{2}$ & $0$ & $\sqrt{6}$ & $\frac{2\sqrt{2}}{3}$ & $0$\\
$\braket{\mathcal{B}_{\pm}^{1,8} \mathcal{B}_{\mp}^{1,8} \mathcal{O}_L}$ & $0$ & $0$ & $\sqrt{6}$ & $\frac{2\sqrt{2}}{3}$ & $0$\\
\hline
$\braket{\mathcal{B}_{\pm}^{1,10} \mathcal{B}_{\pm}^{1,8} \mathcal{O}_L}$ & $-\frac{5+2\sqrt{6}}{\sqrt{2}}$ & $2+\sqrt{6}$ & $\frac{3}{2}$ & $\frac{6\sqrt{2}+\sqrt{3}}{3}$ & $\sqrt{\frac{5}{3}}$\\
$\braket{\mathcal{B}_{\pm}^{1,10} \mathcal{B}_{\mp}^{1,8} \mathcal{O}_L}$ & $\frac{5-2\sqrt{6}}{\sqrt{2}}$ & $2-\sqrt{6}$ & $\frac{3}{2}$ & $\frac{6\sqrt{2}-\sqrt{3}}{3}$ & $\sqrt{\frac{5}{3}}$\\
\hline
$\braket{\mathcal{B}_{\pm}^{1,10} \mathcal{B}_{\pm}^{1,10} \mathcal{O}_L}$ & $-6 \sqrt{2}$ & $4$ & $\sqrt{\frac{3}{2}}$ & $-3\sqrt{2}$ & $3\sqrt{\frac{5}{2}}$ \\
$\braket{\mathcal{B}_{\pm}^{1,10} \mathcal{B}_{\mp}^{1,10} \mathcal{O}_L}$ & $0$ & $4$ & $3\sqrt{\frac{3}{2}}$ & $-\sqrt{2}$ & $\sqrt{\frac{5}{2}}$
\end{tabular}
\caption{Set of three-point correlation functions with operators of different lengths. The length-$L'$ operators $\mathcal{B}_{\pm}^{1,L'}$ are in the twisted sector and each carry two excitations $Y$, whereas $\mathcal{O}_L$ is an untwisted vacuum of length $L$.} \label{tab:Z2CorrelatorsYY}

\begin{tabular}{c||c|c|c|c|c}
Correlators & $L=2$ & $4$ & $6$ & $8$ & $10$\\
\hline \hline
$\braket{\mathcal{B}_{\pm}^{1,8} \hat{\mathcal{B}}_{\pm}^{1,8} \mathcal{O}_L}$ & $0$ & $0$ & $-\frac{1}{\sqrt{6}}$ & $\sqrt{2}$ & $-\frac{\sqrt{5}}{3\sqrt{2}}$ \\
$\braket{\mathcal{B}_{\pm}^{1,8} \hat{\mathcal{B}}_{\mp}^{1,8} \mathcal{O}_L}$ & $0$ & $0$ & $\frac{1}{\sqrt{6}}$ & $\sqrt{2}$ & $\frac{\sqrt{5}}{3\sqrt{2}}$ \\
\hline
$\braket{\mathcal{B}_{\pm}^{1,10} \hat{\mathcal{B}}_{\pm}^{1,8} \mathcal{O}_L}$ & $0$ & $0$ & $0$ & $0$ & $-\sqrt{\frac{5}{2}}$ \\
$\braket{\mathcal{B}_{\pm}^{1,10} \hat{\mathcal{B}}_{\mp}^{1,8} \mathcal{O}_L}$ & $0$ & $0$ & $0$ & $0$ & $\sqrt{\frac{5}{2}}$ \\
\hline
$\braket{\mathcal{B}_{\pm}^{1,10} \hat{\mathcal{B}}_{\pm}^{1,10} \mathcal{O}_L}$ & $0$ & $0$ & $-\frac{\sqrt{3}}{2\sqrt{2}}$ & $\sqrt{2}$ & $0$ \\
$\braket{\mathcal{B}_{\pm}^{1,10} \hat{\mathcal{B}}_{\mp}^{1,10} \mathcal{O}_L}$ & $0$ & $0$ & $\frac{\sqrt{3}}{2\sqrt{2}}$ & $\sqrt{2}$ & $0$
\end{tabular}
\caption{Another set of correlators. The twisted operators $\mathcal{B}_{\pm}^{1,L'}$ carry two excitations $Y$, while the twisted operators $\hat{\mathcal{B}}_{\pm}^{1,L'}$ carry two excitations $\bar{Y}$. Again, $\mathcal{O}_L$ is an untwisted vacuum of length $L$.} \label{tab:Z2CorrelatorsYYB}
\end{table}
Let us begin by considering operators with longitudinal excitations. For this we use two twisted operators as given in Tab. \ref{tab:Z2Operators} carrying two excitations each as well as an untwisted vacuum $\mathcal{O}_L$ of length $L=2,\dots, 10$.
We evaluate these three-point functions using Wick contractions, the spin-chain overlap and the hexagon formalism, finding agreement in all cases considered.
Tab. \ref{tab:Z2CorrelatorsYY} lists the results for both operators carrying the same type of excitation $Y$, while in Tab. \ref{tab:Z2CorrelatorsYYB} the first operator features $Y$ and the second carries excitations $\bar{Y}$. 
The time-consuming step in these calculations is the evaluation using Wick contractions, of which our examples involve up to fifteen.

\paragraph{Transversal excitations.}
\begin{table}[t]
\centering
\begin{tabular}{c||c|c|c|c|c}
Correlators & $L=2$ & $4$ & $6$ & $8$ & $10$\\
\hline \hline
$\braket{\mathcal{C}_{\pm}^{1,6} \hat{\mathcal{C}}_{\pm}^{1,6} \mathcal{O}_L}$ & $-4\sqrt{2}$ & $4$ & $-\sqrt{\frac{3}{2}}$ & $0$ & $0$ \\
$\braket{\mathcal{C}_{\pm}^{1,6} \hat{\mathcal{C}}_{\mp}^{1,6} \mathcal{O}_L}$ & $0$ & $0$ & $\sqrt{\frac{3}{2}}$ & $0$ & $0$ \\
\hline
$\braket{\mathcal{C}_{\pm}^{1,8} \hat{\mathcal{C}}_{\pm}^{1,6} \mathcal{O}_L}$ & $3 \pm 2\sqrt{2}$  & $\mp 3-2\sqrt{3}$ & $\sqrt{3} \pm \sqrt{6}$ & $\mp \sqrt{2}$ & $0$ \\
$\braket{\mathcal{C}_{\pm}^{1,8} \hat{\mathcal{C}}_{\mp}^{1,6} \mathcal{O}_L}$ &  $3 \mp 2\sqrt{2}$ & $\pm 3-2\sqrt{3}$ & $\sqrt{3} \mp \sqrt{6}$ & $-\mp \sqrt{2}$ & $0$ \\
\hline
$\braket{\mathcal{C}_{\pm}^{1,8} \hat{\mathcal{C}}_{\pm}^{1,8} \mathcal{O}_L}$ & $-6\sqrt{2}$ & $8$ & $-5\sqrt{\frac{3}{2}}$ &  $3\sqrt{2}$ & $-\sqrt{\frac{5}{2}}$ \\
$\braket{\mathcal{C}_{\pm}^{1,8} \hat{\mathcal{C}}_{\mp}^{1,8} \mathcal{O}_L}$ & $0$ & $0$ & $\sqrt{\frac{3}{2}}$ & $-\sqrt{2}$  & $\sqrt{\frac{5}{2}}$
\end{tabular}
\caption{Set of correlators with operators carrying transversal excitations. The twisted operators $\mathcal{C}_{\pm}^{1,L}$ carry two excitations $Z$, while the twisted operators $\hat{\mathcal{C}}_{\pm}^{1,L}$ carry two excitations $\bar{Z}$.} \label{tab:Z2CorrelatorsXXB}
\end{table}
The set up here is similar to the case before, though here we use the operators from Tab. \ref{tab:Z2OperatorsX} carrying transversal excitations. 
The splitting factor $\rho^k_{\ell}$ is again given by \eqref{eq:OrbiWeightFactor} with $k_{\ell}=0$.
In Tab. \ref{tab:Z2CorrelatorsXXB} we collected results with the first operator carrying $Z$ and the second operator carrying $\bar{Z}$ excitations. Again we find agreement between tree-level Wick contractions, spin-chain overlaps and the hexagon formalism for all examples considered.

\subsubsection{More general three-point functions}
The generalisation to more complicated correlation functions is straightforward. As an additional example let us consider three non-BPS operators. As before two operators carry orbifold twist and one operator is untwisted. Here we consider the twisted operators with one longitudinal excitation each, \ie $o^{1,L}=\tr \gamma Y X^{L-1}$  which will have momentum $p=\pi$ and rapidity $u=0$. The untwisted operator will be a non-BPS operator of length $L=4$ with two excitations and is given by
\begin{equation}
    \mathcal{B}^{0,4} = \frac{1}{\sqrt{3}} ( \mathcal{O}_0^{0,4} -\mathcal{O}_1^{0,4} )\,.
\end{equation}
Its energy is $E=6$ and the rapidities are $u_1=-u_2=\frac{1}{2\sqrt{3}}$. The results of the correlators consisting of these operators are listed in Tab. \ref{tab:Z2CorrelatorsYYYY}. Again we find agreement between the results from gauge theory and the hexagon formalism. Note that the correlation functions vanish if the lengths of the operators $o^{1,L}$ are not identical.

\begin{table}[t]
\centering
\begin{tabular}{c||c|c|c|c}
Correlators & $L=2$ & $4$ & $6$ & $8$ \\
\hline \hline
$\braket{\mathcal{B}^{0,4} o^{1,L} o^{1,L}}$ & $\sqrt{6}$ & $4\sqrt{\frac{2}{3}}$ & $5\sqrt{\frac{2}{3}}$ & $2\sqrt{6}$\\
\end{tabular}
\caption{A simple example for correlation functions of twisted and untwisted operators all carrying excitations. While the twisted operators $o^{1,L}$ carry one excitation each, the non-BPS operator $\mathcal{B}^{0,4}$ carries two excitations.} \label{tab:Z2CorrelatorsYYYY}
\end{table}

\subsection{The hexagon for general orbifolds} \label{Sec:Further}

Trying to use the adjoint vacuum for the $\amsmathbb{Z}_2$ orbifold turns out to be more challenging. In this Section we discuss the obstacles we face in the construction of translated gauge-theory states. In response to these, we may consider abandoning the translation $\mathcal{T}$ and setting up three-point functions by hand. It is important to discuss these issues extensively here, because for higher-order $\amsmathbb{Z}_M$ orbifolds, the additional $\SU(2)$-symmetry over the bifundamental vacuum is lost and we are faced with the same issues independent of gauge choice.

The first thing to notice is that for the adjoint vacuum the translation $\mathcal{T}$ \eqref{eq:Translation} is a broken generator. Therefore, its action changes the gauge indices of the fields and the algebroid coproduct \eqref{eq:AlgebroidCoproduct} should be used when acting on a multi-particle state. Following the procedure in \cite{Bertle:2024djm}, the trace needs to be cut open and we have to sum over all cut locations. Acting once with the broken generators leads to unphysical intermediate states that cannot close. However, acting multiple times may lead to physical states which are however not necessarily in the same superconformal multiplet as the original state.

Let us illustrate this for the adjoint vacuum of the $\amsmathbb{Z}_2$ orbifold. Recall that we have $(Y,\,Z) \sim (\omega^{-1} Y,\, Z)$. Acting with the  translation on a single field component, we obtain
\begin{equation}\label{eq:trans}
\begin{aligned}
    Z_{11} \quad \to \quad \hat{Z}(t) &= Z_{11} + t (Y_{12} - \bar{Y}_{12}) + t^2 \bar{Z}_{11}\,, \\ 
    Y_{12} \quad \to \quad \hat{Y}(t) &= Y_{12} + t \bar{Z}_{11} \,,
\end{aligned}
\end{equation}
and we see that fields in the adjoint and bifundamental representation mix. Acting with this translation and following the rules given in \cite{Bertle:2024djm}, we project down to the allowed colour index structure only at the end when closing the trace again. Since not all field configurations are allowed, there are no effective propagators \eqref{eq:Propagators} in this case and Wick contractions have to be carried out for individual component fields. Some of the resulting expressions are not even invariant under translation, due to the explicit appearance of $t$ in \eqref{eq:trans}.
We attribute these difficulties to the modified multiplet structure of the orbifold theory. Considering for example a translated twisted vacuum state, we find among many terms some physical configurations carrying two fields $Y$ as
\begin{equation}
    \tr \, \gamma \tilde{Z}(t)^L = \dots{} + t^2 \tr( \cdots Z_{11}Y_{12} Z_{22} \cdots Z_{22} Y_{21} Z_{11}\cdots) + \dots \,.
\end{equation}
Index structure aside, such a state would be allowed as a vacuum descendant in $\mathcal{N}=4$ SYM. Accordingly, the excitations would have vanishing momenta. 
However, for this choice of vacuum in the orbifold theory, there are no descendants that carry two fields $Y$. Solving the Bethe equation $e^{i p_1 (L-1)}=-1$ as $p_2=-p_1$, the solution is $p_1=\pi /(L-1)$ and hence non-vanishing, in contrast with $\mathcal{N}=4$ SYM. In the orbifold theory, this state belongs to a different superconformal multiplet than the vacuum state.

Even though the trace closes under the gauge index structure it seems that the broken translation generates a plethora of states from other multiplets. Let us make this more explicit by considering an example of a twisted vacuum operator of length $L=3$, \ie $\tr \gamma Z^3$. Under the action of the translation, this operator becomes
\begin{equation}
    \tr \gamma \tilde{Z}(t)^3 = \tr\gamma Z^3 + t^2 \left(6  \tr \gamma Y Y Z - 6\tr \gamma Y \bar{Y} Z -  6 \tr \gamma \bar{Y} Y Z + \tr \gamma \bar{Z} Z Z \right) + O(t^4)\,.
\end{equation}
We see that already the term $\tr \gamma Y Y Z$ is a non-BPS state (\cf operator $\mathcal{B}^{1,3}$ in the list of states in Tab. \ref{tab:States} of Appendix \ref{app:Collection}). Similarly the other terms are in general non-BPS, and their integrable description would be governed by more general Bethe equations \cite{Beisert:2005he}.
When we calculate a three-point function involving this operator we would obtain a linear combination of more elementary three-point functions. To match field theory calculations to the hexagon predictions, we therefore need to project out states that are not descendants of the original operator after acting with the translation. We will consider examples for such three-point functions in the following Sec. \ref{Sec:Z2adjLongitudinal}. 

One may now raise the question whether an $R$-rotated translation along the lines of \eqref{eq:Translation} is a sensible operation in this context. If we have to project out essentially all operators, we may as well insert operators by hand and check $R$-charge conservation by inspection. This allows us to compute the individual structure constants and match them to the hexagon predictions. Unsurprisingly, both methods lead to the same results. 

If one wants to salvage the strategy using translations, it is worth noting that the projection to the correct multiplets is to some extent already implemented in the gauge theory calculation. If we strictly move the operators of interest to $t=0$, $1$ and $\infty$, the first one will remain unchanged, the last one will consist purely of $\bar{\Phi}_V$-fields and there will only be certain components of the operator at $t=1$ that can contract nontrivially. Similarly, a restriction to a subset of operators can further reduce the number of possible contractions.

For higher-order orbifolds, a good example of this phenomenon is the $\SU(2)_R$-sector which we may align with transversal excitations over a bifundamental vacuum. The hexagon results agree straightaway with Wick contractions in this case. To illustrate this observation, we collect some exemplary three-point functions for the $\amsmathbb{Z}_3$ orbifold in Sec. \ref{Sec:Z3Examples}. This is similar to the observations made in \cite{Eden:2022ipm} for the $\beta$- and $\gamma$-deformation of $\mathcal{N}=4$ SYM and might be due to a similar projection onto the fields $X$ and $\bar{Y}$ for the three-point function in those cases. 

\subsubsection{Three-point functions in the adjoint gauge} \label{Sec:Z2adjLongitudinal}

\paragraph{Longitudinal excitations.} As discussed above, the obstacle to performing field theory checks is that the translation \eqref{eq:Translation} mixes  different multiplets. We will now consider a specific setup that allows to project down to the three-point function we want to calculate. For this we consider correlators with one twisted non-BPS operator $\mathcal{B}$, one untwisted BPS operator $\mathcal{O}_L$ and one twisted vacuum $\mathcal{O}^{k,L} = \tr \gamma^k Z^L$. Placing these operators at specific points, we can make sure, that we do not generate mixtures of different multiplets.\footnote{Of course, the structure constants will ultimately be independent of the positions of the operators.} For instance, placing the operator $\mathcal{B}$ at the origin, it is not translated at all. We move the twisted vacuum $\mathcal{O}^{k,L}$ to $t=\infty$, which effectively turns all the fields $\tilde{Z}(t)$ into $\bar{Z}$. The untwisted BPS operator $\mathcal{O}_L$ can be inserted at a generic point on the line. The translation mixes the vacuum with its descendents and the only selection rule is that the trace closes. We can now use the three approaches from Sec. \ref{Sec:ThreeApproaches} to evaluate three-point functions.

Let us give some explicit examples for the $\amsmathbb{Z}_2$ orbifold in adjoint gauge. The shortest primary operators of length $L=4,5$ with two excitations are listed in Tab. \ref{tab:BadZ2Operators}.
\begin{table}
\centering
\begin{tabular}{c|c|c|c}
$L$ & Eigenstate & $E$ & $u_1$\\
\hline
$4$ & $ \mathcal{B}^{1,4} = \mathcal{O}^{1,4}_0 $ & 2 & $\frac{\sqrt{3}}{2}$ \\
$5$ & $ \sqrt{-2(2\pm\sqrt{2})} \mathcal{B}^{1,5}_{\pm} = (1\pm \sqrt{2})\mathcal{O}^{1,5}_0 + \mathcal{O}^{1,5}_1 \ $ & $4\mp 2\sqrt{2}$ & $\frac{1}{2} \pm \frac{1}{\sqrt{2}}$
\end{tabular}
\caption{Some examples for twisted $\amsmathbb{Z}_2$-orbifold states of length $L=4,5$ with two excitations on top of the adjoint vacuum. The momentum constraint yields $u_2=-u_1$.} \label{tab:BadZ2Operators}
\end{table}

Using the set-up described above, we can straightforwardly carry out Wick contractions to evaluate three-point functions. The spin-chain overlap is in this case given by
\begin{equation}
    C_Y^{\bullet \circ \circ} = N_1 \delta_{0,K\,\mathrm{mod} M} \sum_{1 \leq n_1 < n_2 \leq \ell_{12}} \frac{\psi^{p_1,p_2}_{n_1,n_2}}{t_{12}^2} \,,
\end{equation}
where we already took into account, that the third operator is placed at $t_3=\infty$. Again, the total twist $K=k_1+k_2+k_3$ of the three-point function has to vanish, which is ensured by the $\delta$-function. Finally, we can evaluate the hexagon form factor finding perfect agreement for all the structure constants considered in Tab. \ref{tab:BadZ2Correlators}.

\begin{table}
\centering
\begin{tabular}{c||c|c|c}
    $\ell_{12}$ & $2$ & $3$ & $4$ \\
    \hline
    $\braket{\mathcal{B}^{1,4} \mathcal{O}_{L} \mathcal{O}^{1,2}}$ 
    & $2$ & $4\sqrt{2}$ & $4\sqrt{3}$   \\
     $\braket{\mathcal{B}^{1,4} \mathcal{O}_{L} \mathcal{O}^{1,3}}$ & $3$ & $2\sqrt{15}$ & $2\sqrt{21}$  \\
     $\braket{\mathcal{B}^{1,4} \mathcal{O}_{L} \mathcal{O}^{1,4}}$ & $4$ & $4\sqrt{6}$ & $8\sqrt{2}$  \\
     \hline
     $\braket{\mathcal{B}^{1,5}_\pm \mathcal{O}_{L} \mathcal{O}^{1,2}}$ & --- & $\sqrt{\frac{3}{2}(10 \pm 7 \sqrt{2})}$ & $\sqrt{5(10 \pm 7 \sqrt{2})}$ \\
     $\braket{\mathcal{B}^{1,5}_\pm \mathcal{O}_{L} \mathcal{O}^{1,3}}$ & $\sqrt{\frac{3}{2}(2 \pm \sqrt{2})}$ & $\sqrt{3(10 \pm 7 \sqrt{2})}$ & $3\sqrt{(10 \pm 7 \sqrt{2})}$ \\
     $\braket{\mathcal{B}^{1,5}_\pm \mathcal{O}_{L} \mathcal{O}^{1,4}}$ &  $\sqrt{3(2 \pm \sqrt{2})}$ & $\sqrt{5(10 \pm 7 \sqrt{2})}$ & $\sqrt{14(10 \pm 7 \sqrt{2})}$
\end{tabular}
\caption{Set of correlators involving two twisted operators and one untwisted operator. Here the adjoint vacuum for the $\amsmathbb{Z}_2$ orbifold is used. Only the twisted operator $\mathcal{B}^{k,L_1}$ carries excitations. The length of the untwisted operator $\mathcal{O}_L$ is for each correlator given as $L=2\ell_{12}-L_1+L_3$.} \label{tab:BadZ2Correlators}
\end{table}

We can generalise the construction of this particularly simple set of observables from $\amsmathbb{Z}_2$ to any $\amsmathbb{Z}_M$. The only difference being that the gauge indices run over a bigger set of numbers. Let us give an explicit example for the $\amsmathbb{Z}_3$ orbifold. Using the adjoint vacuum, the shortes operator we can build is of length $L=4$ carrying three excitations. In fact, the operator takes the simple form
\begin{equation}
    \mathcal{B}^{k,4} = \tr \gamma^k YYYZ \,.
\end{equation}
In the same manner as in Sec. \ref{Sec:Z2adjLongitudinal} we can perform the field theory and compare with the hexagon result. We checked this for correlators with vacuum operators of length $L=2,\dots,6$ and found agreement in all cases, \eg~for $L=6$ we find
\begin{equation}
    \braket{\mathcal{B}^{k,4} \mathcal{O}_6 \mathcal{O}^{-k,4}} = 2\sqrt{6} \,.
\end{equation}

\paragraph{Transversal excitations.} Considering transversal excitations, the eigenstates coincide with the eigenstates given in Tab. \ref{tab:BadZ2Operators}. The calculation of three-point functions is straightforward since we do not need to worry about the mixing of transversal excitations with the vacuum descendants. In particular, we can consider cases in which  only $Z$ and $\bar{Z}$ fields in a translated vacuum state are being contracted. For example, we can calculate the $\amsmathbb{Z}_2$-orbifold correlator
\begin{equation}
    \braket{\mathcal{B}^{1,4}_{X} \mathcal{B}^{1,4}_{\bar{X}} \mathcal{O}_4}  = 2 \,.
\end{equation}

This can easily be extend to other correlators in this sector and even $\amsmathbb{Z}_M$ orbifolds. Using the adjoint vacuum there are however at least $M$ excitations involved, which makes solving the Bethe equations and evaluating the hexagon quite cumbersome.

\subsubsection{$\SU(2)_R$ sector in higher-order orbifolds} \label{Sec:Z3Examples}
In order to present a simple application of the hexagon formalism to higher-order orbifolds, we will now consider a gauge choice that aligns the unbroken $\SU(2)_R$-symmetry with transversal excitations, i.e. $\Phi_V=X$ and $\Phi_T=\bar{Y}$. This corresponds to a different choice for the orientation of $\mathcal{T}$ in \eqref{eq:Translation}. For concreteness, let us consider the scalar sector of the $\amsmathbb{Z}_3$ orbifold. The construction of the spectrum is very similar to the considerations presented in Sec. \ref{sec:Review} and in Sec. \ref{sec:SU2Sectors} for the $\amsmathbb{Z}_2$ orbifold, with the difference that we set $\omega=e^{\frac{2 \pi i}{3}}$ in this case.
We restrict to transversal excitations over the bifundamental vacuum in the $\SU(2)_R$ sector. Recall that the orbifold action on the relevant fields acts as $(X,\,\bar{Y}) \sim (\omega X,\, \omega \bar{Y})$. In order to write the one-loop eigenstates, we will use a short-hand notation as in eq. \eqref{eq:OpBasis}. Due to the index structure of the fields the operators need to be of length $L \text{ mod } 3 = 0$. For the twisted sectors $k=1,2$ we collect the shortest primary states in Tab. \ref{tab:Z3Operators}.
\begin{table}[h]
\centering
\begin{tabular}{c|c|c}
$k$ & Eigenstate & Energy $E$ \\
\hline
$1$ & $\mathcal{B}^{1,6}_1 \sim \left(1-i \sqrt{3}\right) \left(\sqrt{17}-5\right) \mathcal{O}^{1,6}_0 - i \left(\sqrt{3}-i\right) \left(\sqrt{17}-1\right) \mathcal{O}^{1,6}_1 + 4 \mathcal{O}^{1,6}_2$ & $\frac{1}{2} \left(7-\sqrt{17}\right)$ \\
& $\mathcal{B}^{1,6}_2 \sim i \left(\sqrt{3}+i\right) \left(5+\sqrt{17}\right) \mathcal{O}^{1,6}_0 + \left(1+i \sqrt{3}\right) \left(1+\sqrt{17}\right) \mathcal{O}^{1,6}_1 + 4\mathcal{O}^{1,6}_2$ & $\frac{1}{2} \left(7+\sqrt{17}\right)$ \\
$2$ & $\mathcal{B}^{2,6}_1 \sim \left(1+i \sqrt{3}\right) \left(\sqrt{17}-5\right) \mathcal{O}^{2,6}_0 + i \left(\sqrt{3}+i\right) \left(\sqrt{17}-1\right) \mathcal{O}^{2,6}_1 + 4\mathcal{O}^{2,6}_2$  & $\frac{1}{2} \left(7+\sqrt{17}\right)$ \\
 & $\mathcal{B}^{2,6}_2 \sim  i \left(\sqrt{3}-i\right) \left(5+\sqrt{17}\right) \mathcal{O}^{2,6}_0 - \left(1-i \sqrt{3}\right) \left(1+\sqrt{17}\right) \mathcal{O}^{2,6}_1 -4 \mathcal{O}^{2,6}_2$ & $\frac{1}{2} \left(7-\sqrt{17}\right)$
\end{tabular}
\caption{Shortest single-trace operators (not normalised) in the $\SU(2)_R$ sector of the $\amsmathbb{Z}_3$-orbifold theory with length $L=6$ and two excitations. The Bethe eigenstates are given by linear combinations of the basis elements $\mathcal{O}^{k,L}_j$. The operators of the twisted sectors $k=1,2$ are related to each other by complex conjugation of the coefficients.} \label{tab:Z3Operators}
\end{table}
Again, the spectrum can be found from integrability. Using the respective Bethe and momentum equation \eqref{eq:BetheEq}, the corresponding Bethe roots can be worked out. 
The normalised eigenstate can then be obtained using eqs. \eqref{eq:BetheState} and \eqref{eq:BetheNorm}.

As before, we are interested in three-point functions involving two twisted and one untwisted operator. Let us begin with an overlap calculation and assume the following set-up: The operator at position $t=0$ has excitations $\bar{Y}$ on top of the vacuum $X$. At position $t\to\infty$ we have a conjugate operator, namely $Y$ on top of $\bar{X}$. It is clear that the latter state is described by the same Bethe equations with $\omega \to 1/\omega$, as the fields carry opposite charge.
Finally, we have the untwisted vacuum (reservoir) at $t=1$ consisting of fields $\tilde{X}$.

\begin{table}
\centering
\begin{tabular}{c||c|c|c}
    $L$ & $0$ & $6$ & $12$ \\
    \hline
    $\braket{\mathcal{B}^{k,6}_1 \bar{\mathcal{B}}^{k,6}_1 \mathcal{O}_L}$ 
    & $1$ & $0.609612$ & $0$ \\
    $\braket{\mathcal{B}^{k,6}_2 \bar{\mathcal{B}}^{k,6}_3 \mathcal{O}_L}$ 
    & $1$ & $1.64039$ & $0$ \\
    $\braket{\mathcal{B}^{k,6}_1 \bar{\mathcal{B}}^{k,6}_2 \mathcal{O}_L}$ 
    & $0$ & $0$ & $0$
\end{tabular}
\caption{Example of correlators in the $\amsmathbb{Z}_3$ orbifold. The operators $\mathcal{B}_{1,2}^{k,6}$ and $\bar{\mathcal{B}}_{1,2}^{k,6}$ are in the twisted sector with $k=1,2$ and carry two excitations $\bar{Y}$ and $Y$, respectively. The third operator $\mathcal{O}_L$ is a vacuum of length $L$.} \label{tab:Z3CorrelatorsX}
\end{table}

We can evaluate correlators of this form by using Wick contractions (using the effective propagators \eqref{eq:Propagators} irrespective of the gauge index structure involved) or the spin-chain overlap formula from eq. \eqref{eq:OverlapX}. Some simple examples of correlators are given in Tab. \ref{tab:Z3CorrelatorsX}. As discussed above, the excitations $Y,\bar{Y}$ are considered as transversal and therefore only contact-terms contribute in the overlap.
Evaluating the geometric sums of the overlap formula, yields the corresponding hexagon formula for transversal excitations. Hence, the $\SU(2)_R$-sector is governed by the hexagon form factor directly without further admixtures.

It is interesting to note, that we used the conjugate operator at the point $t=\infty$ in this construction. In the original hexagon construction \cite{Basso:2015zoa}, the excitations $X$ would have been placed on the vacuum $Z$ at the origin and then moved by using the translation, \cf the discussion around \eqref{eq:Translation}. However, such an operator cannot exist (not with this length nor in the $\SU(2)_R$ sector). Nonetheless, the hexagon reproduces the correct tree-level result. A similar observation was made in Sec. 4.2 of \cite{Eden:2022ipm} for the $\beta$-deformation of $\mathcal{N}=4$ SYM and even checked at one-loop order.

\section{Conclusions and outlook}\label{Sec:Comclusion}

In this paper we proposed a hexagon formalism for $\mathcal{N}=2$ orbifold theories. It can be seen as a
modification of the $\mathcal{N}=4$ SYM hexagon formalism which takes into account twisted Bethe equations and additional twist factors when taking particles from one hexagon to the other. Although this proposal appears natural from a integrability point of view (e.g. in view of \cite{Beisert:2005he,Skrzypek:2022cgg}), its validity should nevertheless be checked against gauge theory. These tests turn out to be non-trivial already at tree level, due to the breakdown of the $\PSU(2|2)$-symmetry underlying the $\mathcal{N}=4$ case. The lesser issue is posed by the various sectors into which the spectrum of the orbifold theories splits and which have to be taken into account individually. The bigger issue is the breakdown of the $R$-symmetry rotated translation \eqref{eq:Translation} which was introduced in $\mathcal{N}=4$ SYM to preserve $\PSU(2|2)$-symmetry. In the $\mathcal{N}=2$ theories this translation does not commute with the twist and therefore becomes a groupoid object in the language of \cite{Bertle:2024djm}, i.e. it mixes different superconformal multiplets. The hexagon on the other hand only gives individual structure constants, making the matching with gauge theory a daunting task. We nevertheless succeeded in this task for a variety of examples, providing evidence for our proposal.

The operators we considered belong to different $\SU(2)$ sectors built on the bifundamental as well as on the adjoint vacuum introduced in Section \ref{ssec:gauge}. Although presented as a gauge choice, we should add that the operators built on the respective vacua are not related by symmetries. The orbifold action distinguishes them via quiver winding number, twist sector and energy spectrum, so we have to account for them individually. This increases the number of relevant structure constants, making a full catalogue a more challenging endeavour. Our aim was to isolate a few qualitatively distinct examples and to present evidence for the applicability of the hexagon-formalism in these cases. A more complete survey, involving for example twisted-twisted-twisted structure constants or higher-rank sectors is left for future work. 


We considered as instructive example the simplest $\mathcal{N}=2$ $\amsmathbb{Z}_2$-orbifold theory and analysed structure constants of operators built on the bifundamental vacuum. Here an additional $\SU(2)$-symmetry allowed us to perform direct field theory checks at tree-level.
Extending to general $\amsmathbb{Z}_M$ orbifolds the naive application of the $R$-rotated translation \eqref{eq:Translation} spoils a direct field theory check. While the hexagon seems to be inert and reproduces the correct result, translating operators in field theory produces unphysical states and multiplet mixing. After a careful projection on the correct multiplets we find perfect agreement. Alongside these consistency checks, we explicitly computed various structure constants in the $\amsmathbb{Z}_2$- and $\amsmathbb{Z}_3$-orbifold theories, which are new results in their own right.

The results presented in this article were evaluated at tree-level only. Obtaining the asymptotic (in large operator lengths $L$) result is quite straightforward. At a certain order in the coupling $g$, we can solve the Bethe equations perturbatively and expand eq. \eqref{eq:Hex3ptPartition} to that order. 
Finite-size effects are suppressed by the edge width $\ell$ as they will start to contribute at order $g^{2(\ell+1)}$. In order to account for these, a full set of states has to be inserted on the virtual edges \cite{Basso:2015zoa}.
In \cite{Ferrando:2025qkr} the evaluation of gluing and wrapping contributions in orbifold theories were discussed. Considering three-point functions of twisted BPS operators, the known result from localisation \cite{Billo:2022xas,Billo:2022gmq,Billo:2022fnb} was recovered. Following along these lines it would be interesting to combine the investigation of non-BPS operators initiated here with the gluing prescription from \cite{Ferrando:2025qkr} and check their consistency with more advanced gauge-theory results. However, due to the presence of the excitations, the evaluation of even the first correction at gluing order is quite involved for the correlators presented here. We leave this problem for future work but are optimistic that a full determination of the structure constants of orbifold theories at finite coupling may be within reach.

\begin{figure}[h]
    \centering
    \includegraphics[width = .8\textwidth]{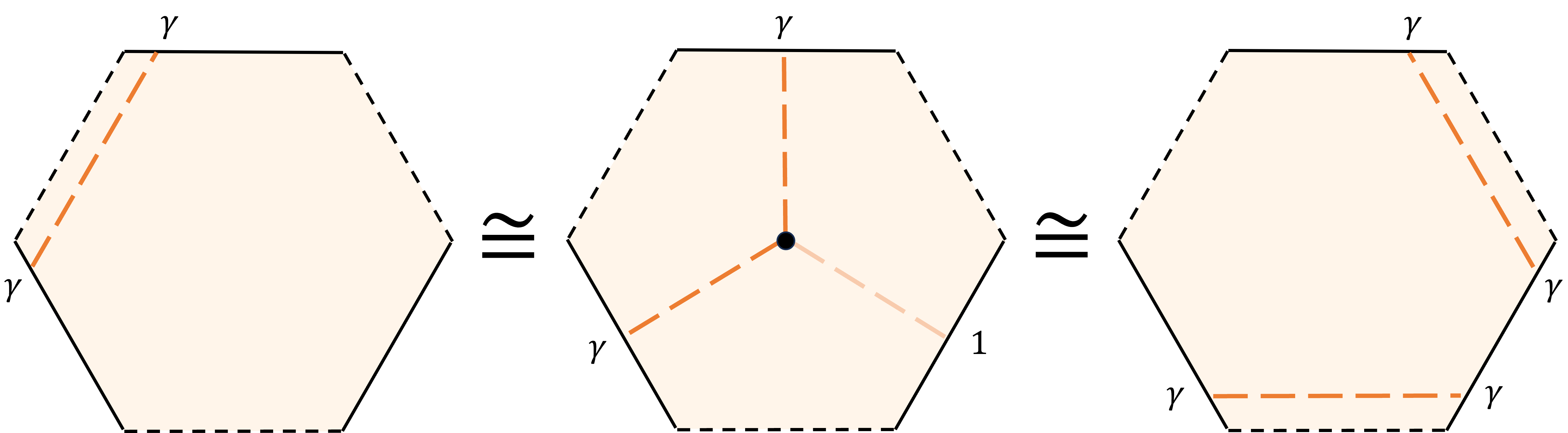}
    \caption{Naive equivalence of twist line distributions on the $\amsmathbb{Z}_2$-orbifolded hexagon. At tree level this equivalence holds, since only the overall twist on each edge is significant. However, when introducing virtual particles for glueing and wrapping effects, this equivalence may come into question.}
    \label{Fig:Equivalence}
\end{figure}

One intriguing qualitative question\footnote{We thank the referee for raising this point.} in this endeavour concerns the placement of the twist operator in Fig. \ref{Fig:TwistHexagon}. Our formalism naturally allows for the twist operators to be moved around the twisted-sector states, as explained in Section \ref{ssec:hexagon}. Pictorially, one could also imagine moving twist lines around the hexagons while keeping their ends fixed. Additionally we may insert operators $\mathbb{1}=\gamma\gamma^{-1}$ and merge twist lines. One may then ask whether all such twist line configurations should be considered as equivalent. A salient example is depicted in Fig. \ref{Fig:Equivalence}. In \cite{Ferrando:2025qkr}, two individual twist lines were introduced which connect two twisted sector states to the same untwisted sector state, their respective twists cancelling in the latter (right-hand side in Fig. \ref{Fig:Equivalence}). From a tree-level perspective this set-up is equivalent to our single line in Fig. \ref{Fig:TwistHexagon} (left-hand side in Fig. \ref{Fig:Equivalence}), but this equivalence is not obvious once virtual particles are involved. Suffice to say that a final verdict on the mobility of twist lines on the hexagon is still to be rendered.

The hexagon formalism in $\mathcal{N}=4$ SYM also paved a way towards the computation of higher-point functions \cite{Eden:2016xvg,Fleury:2016ykk} and non-planar corrections \cite{Eden:2017ozn,Bargheer:2017nne,Bargheer:2019kxb}. In orbifold theories a similar extension may be possible if we meticulously keep track of the twist operators involved. A major obstacle to such considerations is the $R$-rotated translation $\mathcal{T}$ \eqref{eq:Translation}, which plays a central role in the setup of higher-point functions, since we now require an explicit space-time dependence. It is therefore important to gain sufficient control over the algebroid structure generated by $\mathcal{T}$ (\cf\cite{Bertle:2024djm}) or to develop an alternative approach without $R$-rotation.

Apart from $\mathcal{N}=2$-supersymmetry preserving orbifold theories, one could also attempt to extend the hexagon formalism to $\mathcal{N}=1$ and $\mathcal{N}=0$ orbifolds. Although the required technology should be analogous to the one discussed in this paper, the physical interpretations may become more challenging, especially in the latter case where tachyons may appear in the spectrum \cite{Polyakov:1998ju,Klebanov:1998yya,Klebanov:1999ch,Pomoni:2009joh,Skrzypek:2021eue}. Insight from orbifold structure constants may also provide inspiration for the further development of structure constant in the $\beta$- and $\gamma$-deformed theories \cite{Eden:2022ipm}. Finally, one could consider marginal deformations of $\mathcal{N}=2$ orbifold theories \cite{Gadde:2009dj,Pomoni:2021pbj,Bozkurt:2024tpz,Bozkurt:2025exl}, which break conventional integrability but might still allow for a treatment of structure constants in the language of \cite{Bertle:2024djm}.

Let us close our conclusion by calling attention to an interesting discussion of colour-twist operators considered in \cite{Cavaglia:2020hdb}, which we would like to make contact to in the future.

\section*{Acknowledgements} 
We thank Zoltan Bajnok, Burkhard Eden, Simon Ekhammar, Paul Ryan, Alessandro Sfondrini and Konstantinos Zoubos for discussions on many points related to this work. We thank Alessandro Pini and Paolo Vallarino for clarifying comments on related localisation results. DlP is grateful for the funding through a Feodor Lynen Fellowship by the Alexander von Humboldt Foundation. Furthermore, DlP acknowledges partial support by NKFIH Grant K134946. TS would like to thank the Cluster of Excellence EXC 2121 Quantum Universe 390833306 and the Collaborative Research
Center SFB1624 for creating a productive research environment at DESY.

\appendix
\section{Collection of twisted-sector $\SU(2)$ states}\label{app:Collection}

As discussed in Sec. \ref{sec:Review}, we can either choose an adjoint (uncharged) or a bifundamental (charged) vacuum to build our excited states on. 
Choosing the bifundamental vacuum, only excited states can exist in the twisted sector. The twisted vacuum state vanishes as can easily be seen by writing the fields with their gauge group indices $\tr (Z_{12}Z_{21})^{L/2}-\tr (Z_{21}Z_{12})^{L/2} = 0$. 
For the reader's convenience we now collect the twisted-sector $\amsmathbb{Z}_2$-orbifold states up to length $L=10$ with one and two magnons in all the different $\SU(2)$ sectors.
For the reader's convenience we now collect the twisted-sector $\amsmathbb{Z}_2$-orbifold states up to length $L=10$ with one and two magnons in all the different $\SU(2)$ sectors.

Operators with one excitation can only exist over the bifundamental vacuum. The momentum of the magnon is $p=\pi$ (with rapidity $u=0$) and the single-trace operator is given as $\tr \gamma \Phi_E X^{L-1}$ with the energy $E=4$. A single-excitation state over the adjoint vacuum cannot exist, as the gauge indices do not close under the trace.
The list of operators with two excitations is given in Tab. \ref{tab:States}.
It is necessary to distinguish whether the vacuum is charged or uncharged. For a charged vacuum (bifundamental vacuum) carrying charged excitations we have the $\SU(2)_L$ and $\SU(2)_R$ sectors, which are degenerate in the $\amsmathbb{Z}_2$ case. If the excitations are uncharged we are in a mixed $\SU(2)$ sector.

For the uncharged vacuum (adjoint vacuum) the $\SU(2)$ symmetry is always mixed as all bosonic excitations are charged.
\flushbottom
{ \begin{landscape}
\begin{minipage}{0.98\linewidth}
    \noindent
\resizebox{\linewidth}{!}{
\begin{tabular}{c|c|c||c|c|c|c}
    $\amsmathbb{Z}_2$ vac. & Sector & Length & Eigenstate & Energy & $u_1$ & $u_2$ \\
    \hline \hline
    bif. & $\SU(2)_{L,R}$ & $l\in 2\amsmathbb{N}$ & $  \mathcal{B}^{1,l}_{\circ} = \mathcal{O}^{1,l}_0$ & 2 & $\pm \frac{i}{2}$ & $\mp \frac{i}{2}$ \\
    & & $8$ & $\sqrt{6}  \mathcal{B}^{1,8}_{\pm} = \mathcal{O}^{1,8}_1 \pm i \sqrt{3}\mathcal{O}^{1,8}_2 - \mathcal{O}^{1,8}_3$ & 4 & $\pm \frac{\sqrt{3}}{2}$ & $\pm \frac{1}{2\sqrt{3}}$ \\
    & & $10$ & $2 \mathcal{B}^{1,10}_{\pm} = \mathcal{O}^{1,10}_1 \pm i \sqrt{2}\mathcal{O}^{1,10}_2 - \mathcal{O}^{1,10}_3$ & 4 & $\frac{1}{2} \pm \frac{1}{\sqrt{2}}$ &$ -\frac{1}{2} \pm \frac{1}{\sqrt{2}}$  \\
    \hline 
    & mixed $\SU(2)$ & $l\in 2\amsmathbb{N}$ & $2  \mathcal{C}^{1,l}_{\circ} = \mathcal{O}^{1,l}_0$ & 2 & $\pm \frac{i}{2}$ & $\mp \frac{i}{2} $ \\
    &  & $6$ & $2  \mathcal{C}^{1,6}_{\pm} = \sqrt{2} \mathcal{O}^{1,6}_1 \pm i \mathcal{O}^{1,6}_2$ & 4 & $\frac{1}{2} \pm \frac{1}{\sqrt{2}}$ & $-\frac{1}{2} \pm \frac{1}{\sqrt{2}}$ \\
    & & $8$ & $\sqrt{2}  \mathcal{C}^{1,8}_{\pm} = \mathcal{O}^{1,10}_1 \pm i \mathcal{O}^{1,10}_2$ & 4 & $\pm 1 + \frac{\sqrt{3}}{2}$ &$ \pm 1 - \frac{\sqrt{3}}{2}$  \\
    & & $10$ & $ \mathcal{C}_{1\pm}^{1,10}=-0.270598 \mathcal{O}^{1,8}_1 \mp \frac{1}{2} i\mathcal{O}^{1,8}_2 + 0.653281 \mathcal{O}^{1,8}_3 \pm 0.353553i \mathcal{O}^{1,8}_4$ & $4$ & $\pm 0.334089$ & $\pm 0.748303$ \\
    & & & $ \mathcal{C}_{2\pm}^{1,10}=- 0.653281 \mathcal{O}^{1,8}_1 \mp \frac{1}{2} i\mathcal{O}^{1,8}_2 - 0.270598 \mathcal{O}^{1,8}_3 \mp 0.353553i \mathcal{O}^{1,8}_4$ &$4$ & $\pm 2.51367$ & $\pm 0.0994562$ \\

    \hline \hline
    adj. & mixed $\SU(2)$ & $3$ & $ \mathcal{B}^{1,3} = \mathcal{O}^{1,3}_0 $ & 4 & $\frac{1}{2}$ & $-u_1$ \\
    & & $4$ & $ \mathcal{B}^{1,4} = \mathcal{O}^{1,4}_0 $ & 2 & $\frac{\sqrt{3}}{2}$ & $-u_1$ \\
    & & $5$ & $ \sqrt{-2(2\pm\sqrt{2})}  \mathcal{B}^{1,5}_{\pm} = (1\pm \sqrt{2})\mathcal{O}^{1,5}_0 + \mathcal{O}^{1,5}_1 \ $ & $4\mp2\sqrt{2}$ & $\frac{1}{2} \pm \frac{1}{\sqrt{2}}$ & $-u_1$   \\
    & & $6$ & $ \mathcal{B}^{1,6}_{+} =0.850651 \mathcal{O}^{1,6}_0 +0.525731 \mathcal{O}^{1,6}_1$ & $3-\sqrt{5}$ & $\frac{1}{2} \sqrt{5+2\sqrt{5}}$ & $-u_1$ \\
    & & & $ \mathcal{B}^{1,6}_{-} =0.525731 \mathcal{O}^{1,6}_0 -0.850651 \mathcal{O}^{1,6}_1$ & $3+\sqrt{5}$ & $\frac{1}{2} \sqrt{5-2\sqrt{5}}$ & $-u_1$ \\
    & & $7$ & $\sqrt{3} \mathcal{B}^{1,7}_{1} = (\mathcal{O}^{1,7}_0 - \mathcal{O}^{1,7}_1 - \mathcal{O}^{1,7}_2)$ & $4$ & $\frac{1}{2}$ & $-u_1$ \\
    & & & $ \mathcal{B}^{1,7}_{+} =0.788675 \mathcal{O}^{1,7}_0 +0.57735 \mathcal{O}^{1,7}_1 +0.211325 \mathcal{O}^{1,7}_2$ & $4-2\sqrt{3}$ & $1+ \frac{\sqrt{3}}{2}$ & $-u_1$ \\
    & & & $ \mathcal{B}^{1,7}_{-} =0.211325 \mathcal{O}^{1,7}_0 -0.57735 \mathcal{O}^{1,7}_1 +0.788675 \mathcal{O}^{1,7}_2$ & $4+2\sqrt{3}$ & $1- \frac{\sqrt{3}}{2}$ & $-u_1$ \\
    & & $8$ & $ \mathcal{B}^{1,8}_{1} = 0.327985 \mathcal{O}^{1,8}_0 -0.736976 \mathcal{O}^{1,8}_1 +0.591009 \mathcal{O}^{1,8}_2$ & $6.49396$ & $0.240787$ & $-u_1$ \\
    & & & $ \mathcal{B}^{1,8}_{2} = 0.591009
   \mathcal{O}^{1,8}_0 -0.327985 \mathcal{O}^{1,8}_1 -0.736976 \mathcal{O}^{1,8}_2$ & $3.10992$ & $0.62698$ & $-u_1$ \\
    & & & $ \mathcal{B}^{1,8}_{3} = 0.736976 \mathcal{O}^{1,8}_0 +0.591009 \mathcal{O}^{1,8}_1 +0.327985 \mathcal{O}^{1,8}_2$ & $0.396125$ & $2.19064$ & $-u_1$ \\
    & & $9$ & $ \mathcal{B}^{1,9}_{1} = 0.392847 \mathcal{O}^{1,8}_0 -0.69352 \mathcal{O}^{1,8}_1 +0.13795 \mathcal{O}^{1,8}_2 +0.587938
   \mathcal{O}^{1,8}_3$ & $5.53073$ & $0.334089$ & $-u_1$ \\
   & & & $\mathcal{B}^{1,9}_{2} = 0.69352 \mathcal{O}^{1,8}_0 +0.587938 \mathcal{O}^{1,8}_1 +0.392847
   \mathcal{O}^{1,8}_2 +0.13795 \mathcal{O}^{1,8}_3$ & $0.304482$ & $2.51367$ & $-u_1$ \\
   & & & $\mathcal{B}^{1,9}_{3} = 0.13795 \mathcal{O}^{1,8}_0 -0.392847
   \mathcal{O}^{1,8}_1 +0.587938 \mathcal{O}^{1,8}_2 -0.69352\mathcal{O}^{1,8}_3$ & $7.69552$ & $0.0994562$ & $-u_1$ \\
   & & & $\mathcal{B}^{1,9}_{4} = 0.587938
   \mathcal{O}^{1,8}_0 -0.13795 \mathcal{O}^{1,8}_1 -0.69352 \mathcal{O}^{1,8}_2 -0.392847
   \mathcal{O}^{1,8}_3$ & $2.46927$ & $0.748303$ & $-u_1$ \\
   & & $10$ & $\mathcal{B}^{1,10}_{1} = 0.57735 (\mathcal{O}^{1,10}_0-\mathcal{O}^{1,10}_2-\mathcal{O}^{1,10}_3)$ & $2$ & $\frac{\sqrt{3}}{2}$ & $-u_1$ \\
   & & & $\mathcal{B}^{1,10}_{2} = 0.228013
   \mathcal{O}^{1,10}_0 -0.57735 \mathcal{O}^{1,10}_1 +0.656539 \mathcal{O}^{1,10}_2 -0.428525
   \mathcal{O}^{1,10}_3$ & $7.06418$ & $0.181985$ & $-u_1$ \\
   & & & $\mathcal{B}^{1,10}_{3} = 0.428525 \mathcal{O}^{1,10}_0 -0.57735 \mathcal{O}^{1,10}_1 -0.228013
   \mathcal{O}^{1,10}_2 +0.656539 \mathcal{O}^{1,10}_3$ & $4.69459$ & $0.41955$ & $-u_1$ \\
   & & & $\mathcal{B}^{1,10}_{4} = 0.656539 \mathcal{O}^{1,10}_0 +0.57735
   \mathcal{O}^{1,10}_1 +0.428525 \mathcal{O}^{1,10}_2 +0.228013 \mathcal{O}^{1,10}_3$ & $0.24123$ & $2.83564$ & $-u_1$ \\
\end{tabular}}

\captionof{table}{List of operators with two bosonic excitations and up to length $L=10$ for $\amsmathbb{Z}_2$ orbifolds. We need to distinguish the different sectors as well as the vacua on which they are build.}\label{tab:States}
\end{minipage}
\end{landscape}
}

\section{Contact terms} \label{App:ContactTerms}
Let us spell out the overlap formula for two operators carrying two longitudinal excitations $\Phi_L$ each. The overlap for $C^{\bullet \bullet \circ}_{\Phi_L,\Phi_L}$ can be expressed by removing the contact terms from $C^{\bullet \bullet \circ}_{\Phi_L,\bar{\Phi}_L}$ in \eqref{eq:OverlapYYB}. Explicitly, it is given by
\begin{equation}
    \begin{aligned}
C^{\bullet \bullet \circ}_{\Phi_L,\Phi_L} = C^{\bullet \bullet \circ}_{\Phi_L,\bar{\Phi}_L} - N_1 N_2 \bigg( 
&\sum_{\substack{1 \leq n_1 < n_2 \leq \ell_{12} \\ \ell_{23} < m_1 \leq L_2 - n_1}}
\psi^{p_1,p_2}_{n_1, n_2} \, \psi^{p_3,p_4}_{m_1, L_2 - n_1 + 1}
\cdot \frac{1}{(t_1 - t_2)^2} \cdot \frac{1}{(t_2 - t_1)^2} \\
+ &\sum_{\substack{1 \leq n_1 < \ell_{12} \\ \ell_{12} < n_2 \leq L_1 \\ \ell_{23} < m_1 \leq L_2 - n_1}}
\psi^{p_1,p_2}_{n_1, n_2} \, \psi^{p_3,p_4}_{m_1, L_2 - n_1 + 1}
\cdot \frac{-1}{(t_1 - t_2)^2} \cdot \frac{1}{(t_1 - t_3)} \cdot \frac{1}{(t_2 - t_1)} \\
+ &\sum_{\substack{1 \leq n_1 < n_2 \leq \ell_{12} \\ 1 \leq m_1 \leq \ell_{23} }}
\psi^{p_1,p_2}_{n_1, n_2} \, \psi^{p_3,p_4}_{m_1, L_2 - n_1 + 1}
\cdot \frac{-1}{(t_1 - t_2)^2} \cdot \frac{1}{(t_1 - t_2)} \cdot \frac{1}{(t_2 - t_3)} \\
+ &\sum_{\substack{1 \leq n_1 \leq \ell_{12} < n_2 \leq L_1 \\ 1 \leq m_1 \leq \ell_{23} }}
\psi^{p_1,p_2}_{n_1, n_2} \, \psi^{p_3,p_4}_{m_1, L_2 - n_1 + 1}
\cdot \frac{-1}{(t_1 - t_2)} \cdot \frac{1}{(t_1 - t_3)} \cdot \frac{1}{(t_2 - t_3)} \\
+ &\sum_{\substack{1 < n_1 < n_2 \leq \ell_{12}\\ L_2- n_1 + 1 < m_2 \leq L_2 }}
\psi^{p_1,p_2}_{n_1, n_2} \, \psi^{p_3,p_4}_{L_2 - n_1 + 1, m_2}
\cdot \frac{1}{(t_1 - t_2)^2} \cdot \frac{1}{(t_2 - t_1)^2} \\
+ &\sum_{\substack{1 < n_1 \leq \ell_{12} < n_2 \leq L_1 \\ L_2-n_1+1 < m_2 \leq L_2}}
\psi^{p_1,p_2}_{n_1, n_2} \, \psi^{p_3,p_4}_{L_2 - n_1 + 1, m_2}
\cdot \frac{-1}{(t_1 - t_2)^2} \cdot \frac{1}{(t_2 - t_1)} \cdot \frac{1}{(t_1 - t_3)} \\
+ &\sum_{\substack{1 \leq n_1 < n_2 \leq \ell_{12} -1 \\ \ell_{23} < m_1 \leq L_2-n_2}}
\psi^{p_1,p_2}_{n_1, n_2} \, \psi^{p_3,p_4}_{m_1, L_2 - n_2 + 1}
\cdot \frac{1}{(t_1 - t_2)^2} \cdot \frac{1}{(t_2 - t_1)^2} \\
+ &\sum_{\substack{1 \leq n_1 < n_2 \leq \ell_{12} \\ 1\leq m_1 \leq \ell{23} }}
\psi^{p_1,p_2}_{n_1, n_2} \, \psi^{p_3,p_4}_{m_1, L_2 - n_2 + 1}
\cdot \frac{-1}{(t_1 - t_2)^2} \cdot \frac{1}{(t_1 - t_2)} \cdot \frac{1}{(t_2 - t_3)} \\
+ &\sum_{\substack{1 \leq n_1 < n_2 \leq \ell_{12}  \\ L_2-n_2+1 < m_2 \leq L_2}}^{}
\psi^{p_1,p_2}_{n_1, n_2} \, \psi^{p_3,p_4}_{L_2 - n_2 + 1, m_2}
\cdot \frac{1}{(t_1 - t_2)^2} \cdot \frac{1}{(t_2 - t_1)^2} \\
- &\sum_{\substack{1 \leq n_1 < n_2 \leq \ell_{12} }}
\psi^{p_1,p_2}_{n_1, n_2} \, \psi^{p_3,p_4}_{L_2 - n_2 + 1, L_2 - n_1 + 1}
\cdot \frac{1}{(t_1 - t_2)^2} \cdot \frac{1}{(t_2 - t_1)^2} \bigg)\,.
\end{aligned}
\end{equation}

\bibliography{refs}
\end{document}